\documentclass[10pt,twocolumn,letterpaper]{article}

\usepackage[pagenumbers]{cvpr} 

\usepackage{placeins}
\usepackage{cuted}

\definecolor{saffron}{rgb}{0.95,0.75,0.2}

\makeatletter
\renewcommand{\paragraph}{%
  \@startsection{paragraph}{4}%
  {\z@}{-0.5em}{-0.5em}%
  {\normalfont\normalsize\bfseries}%
}
\makeatother

\definecolor{cvprblue}{rgb}{0.21,0.49,0.74}
\usepackage[pagebackref,breaklinks,colorlinks,allcolors=cvprblue]{hyperref}

\def\confName{CVPR}
\def\confYear{2027}

\title{
Read My Lips: Context-aware Lip Reading Using Language\\and Audio-Visual Models
}
\author{
Amir Razlighi$^{1,2,3}$\qquad Zhuo Ning$^{1}$\qquad Naomi Ken Korem$^{4}$\qquad Nisarg Patel$^{5}$\\[3pt]
Yiorgos Chrysanthou$^{2,3}$\qquad Daniel Cohen-Or$^{6}$\qquad Ali Mahdavi Amiri$^{1}$\\[9pt]
{\small $^{1}$Simon Fraser University\hspace{0.65em}$^{2}$University of Cyprus\hspace{0.65em}$^{3}$CYENS Centre of Excellence\hspace{0.65em}$^{4}$Lightricks\hspace{0.65em}$^{5}$MARZ VFX\hspace{0.65em}$^{6}$Tel Aviv University}
}

\begin{document}
\twocolumn[{%
\renewcommand\twocolumn[1][]{#1}%
\maketitle
\vspace{-1.5em}   
\begin{center}
    \centering
    \includegraphics[width=\textwidth]{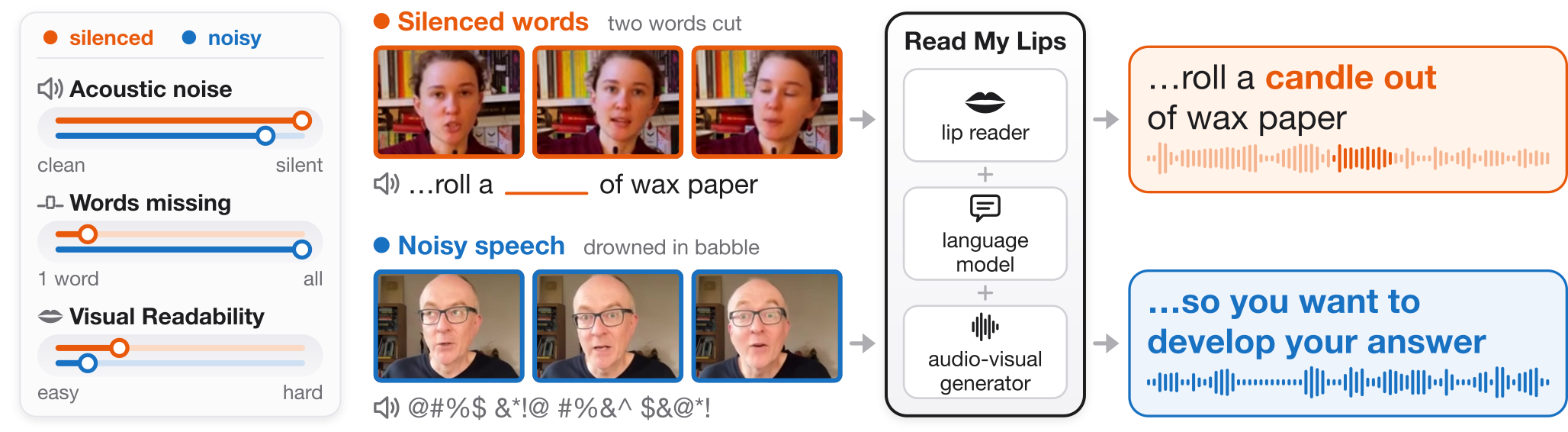}
    \captionof{figure}{\textbf{Read My Lips recovers a variety of corruptions in audio} by using a combination of a lip reader, a language model, and an audio-visual generator. \emph{Left:} the inputs vary in how damaged the audio is, how many words are affected (missing), and how hard the lips are to read. \emph{Right:} two examples with different corruptions. When two words are cut from the audio ({\color{orange}orange}), Read My Lips fills them in; when the whole clip is drowned in babble noise ({\color{blue}blue}), it recovers the sentence from the lips.}
    \label{fig:teaser}
\end{center}%
}]

\begin{abstract}
Real-world speech is often partially observable: audio may be missing or corrupted by noise and competing speakers, while the speaker's lips remain visible. Recovering speech requires combining what can be heard, seen, and inferred from context. We introduce RML (\emph{Read My Lips}), a collaborative method combining a lip reader, a language model, and an audio-visual generative model. The lip reader provides articulation hypotheses, the language model uses context to resolve ambiguous or missing words, and the audio-visual generator reconstructs synchronized speech while preserving speaker identity. Since source reliability varies, RML generates multiple reconstructions and uses the available evidence to favor the most reliable ones. This allows RML to handle missing audio, strong interference, and poor visual evidence without manually defined thresholds. Quantitative and qualitative experiments across these conditions show the benefit of combining visual, acoustic, and linguistic signals.
\end{abstract}
    
\section{Introduction}

Imagine trying to follow a conversation at a crowded reception. Several people are speaking at once, music and background noise overwhelm parts of the speech, yet the person in front of you is clearly visible. Even when their voice becomes difficult to hear, their lip movements and the context of the conversation still provide clues about what they are saying. Similar situations arise in video calls with dropped audio, recordings made in noisy environments, archival footage with damaged sound, and film or content production where corrupted speech needs to be restored. In all these cases, recovering speech requires using whatever information remains available across the audio, video, and surrounding conversation.

More generally, real-world speech is often only partially observable. Background noise or competing speakers may obscure the audio, words may disappear due to recording or transmission failures, and the speaker's mouth may be occluded or seen from a difficult viewpoint. We characterize these challenges along three dimensions, illustrated in Fig.~\ref{fig:teaser}: \emph{acoustic noise}, from clean speech to severe noise or competing speakers; \emph{missing words}, from isolated words to large portions of an utterance; and \emph{visual readability}, from a clear frontal face to difficult viewpoints, occlusions, or animated and generated speakers. These dimensions can vary independently, creating conditions in which no single modality is consistently reliable.

Recovering speech under these conditions requires combining three complementary sources of evidence. \emph{Visual articulation} provides information about the spoken sounds, but is inherently ambiguous because different phonemes can produce similar lip movements. \emph{Acoustic evidence}, even when heavily corrupted, can provide cues about phonemes, timing, and speaker identity. Finally, \emph{linguistic context} can resolve ambiguities using the surrounding reliable speech. For example, in ``I ordered a --- for lunch,'' context strongly constrains the missing word. However, context alone is unreliable for names, numbers, or unexpected expressions. Effective restoration must therefore combine what can be \emph{seen}, \emph{heard}, and \emph{inferred}.

Existing models provide complementary capabilities for these sources. Lip-reading models capture fine-grained visual articulation but struggle with ambiguous visemes, occlusion, and difficult viewpoints. Large language models use linguistic context to resolve ambiguities, but may favor plausible words over what was actually articulated. Audio-visual generative models can synthesize synchronized and speaker-consistent speech, but are not specialized for precise lip reading. Their complementary strengths motivate combining them rather than relying on any one model.

We introduce RML (\emph{Read My Lips}), a collaborative framework for audio-visual speech restoration. Given a video and potentially corrupted audio, a \emph{lip reader} provides hypotheses about what is articulated, a \emph{language model} reasons about what is plausible given the surrounding speech, and an \emph{audio-visual generative model} reconstructs the final speech using these predictions together with the video and remaining acoustic evidence. Their relative importance naturally changes with the input: visual evidence dominates when audio is missing, context helps resolve ambiguous articulation, and remaining audio constrains pronunciation, timing, and speaker identity.

RML thus formulates speech restoration as a multimodal inference problem (see Figure~\ref{fig:teaser}): the lip reader estimates \emph{what appears to be articulated}, the language model reasons about \emph{what is plausible in context}, and the audio-visual model determines \emph{how the recovered speech should sound and align with the video}. 
To adapt to different input conditions, RML generates multiple reconstructions using different component and audio conditions, scores their support from the observed audio and lip motion, and combines them through evidence-weighted minimum Bayes risk to favor the most reliable reconstructions. We evaluate RML across different levels of acoustic corruption, missing speech, and visual readability, demonstrating the benefit of combining these complementary sources of information.

\section{Background: Phonemes and Visemes}

Phonemes are the basic sound units that distinguish words in a spoken
language. For example, /p/, /b/, and /m/ are different phonemes.
Visemes are their visual counterparts: the mouth and lip configurations
seen when a person speaks. However, there is no one-to-one correspondence
between phonemes and visemes. Different sounds can produce nearly
identical mouth movements. For example, /p/, /b/, and /m/ all involve
closing the lips, making words such as \emph{pat}, \emph{bat}, and
\emph{mat} difficult to distinguish visually. This ambiguity makes
lip reading challenging even when the mouth is clearly visible.

Additional information can help resolve these ambiguities. Linguistic
context constrains which words are likely in a sentence, while even
weak acoustic cues may distinguish sounds that look the same on the
lips. This explains why combining visual, acoustic, and linguistic
evidence can be more effective than relying on lip reading alone.

\section{Related Work}
\label{sec:related}

\paragraph{Visual speech recognition.}
Lip reading has advanced through audio-visual self-supervised pretraining~\cite{shi2022learning} and through training on large automatically labeled or pseudo-labeled corpora~\cite{ma2023auto,NEURIPS2024_usr,haliassos2026pay}. Native readers now also exist for languages other than English~\cite{Ma2022}. A recent line of work connects visual features to large language models. Some methods feed visual speech units or features directly into the LLM~\cite{yeo2024visual,10889251}; others predict phonemes that the LLM then assembles into a sentence~\cite{Thomas_2025_ICCV}. These models draw on the LLM's knowledge of language. Used as lip readers, however, they see only the video and ignore the speech that is still audible around the words they must read. We use a lip reader as a black box that returns an $N$-best list, and bring in the audible context at the next stage.

\paragraph{Language models for recognition error correction.}
Generative error correction trains an LLM to map the $N$-best list of a speech recognizer to a corrected transcript~\cite{NEURIPS2023_64922674}. Audio-visual variants condition this correction on lip features~\cite{ghosh24b_interspeech} or let the LLM combine separate audio and visual $N$-best lists~\cite{kim-etal-2026-two}. Our language model also combines a lip reader’s hypotheses with a transcript of the audio, but it decides what was said in a silenced span from the speech around it and the span’s duration.

\begin{figure*}[t]
    \centering
    \includegraphics[width=\linewidth]{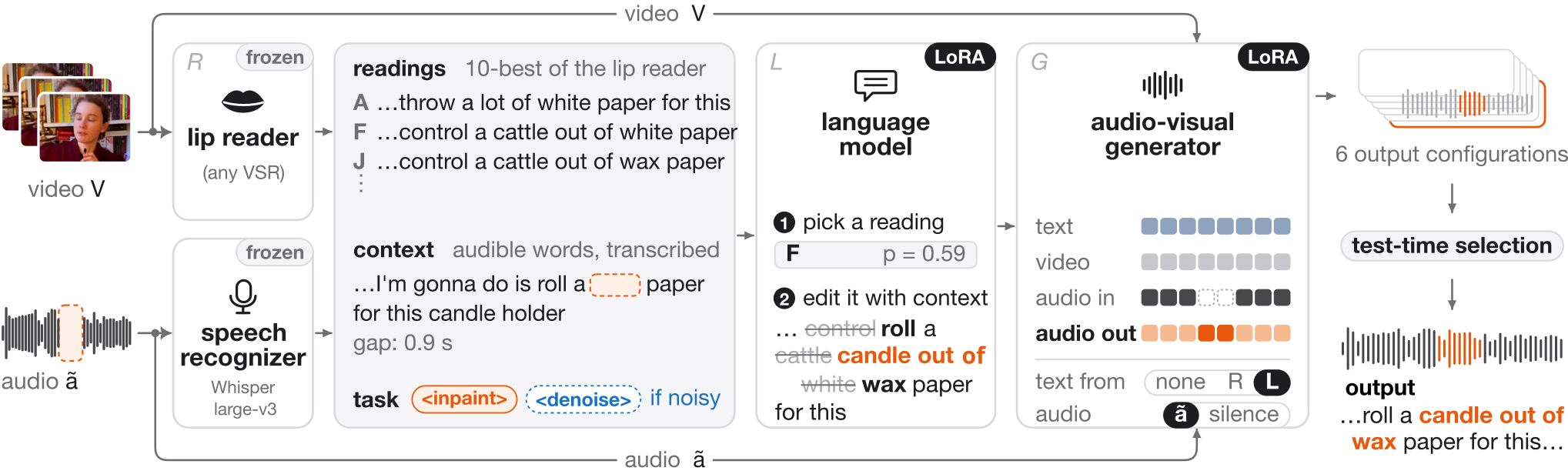}
    \caption{\textbf{Overview of Read My Lips} on a real test clip with silenced words.
  A frozen lip reader $R$ returns its $N$-best readings of the whole clip, and a frozen speech recognizer
  transcribes the audible context (if present). The language model $L$ picks the closest reading and edits it to fit the
  context. The audio-visual generator $G$ receives the video, an audio track and the text as clean in-context
  tokens and generates the speech. Switching the text source (none, $R$, or $L$) and the audio track
  ($\tilde{a}$ or silence) gives six possible configurations for $G$; a test-time selection (\cref{sec:method:selection}) chooses the best configuration.}
  \label{fig:pipeline}
\end{figure*}

\paragraph{Video-to-speech synthesis.}
Video-to-speech methods generate speech from a silent video. Early work trained one model per speaker~\cite{Prajwal_2020_CVPR}; later methods work in the wild, through intermediate speech units~\cite{choi23_interspeech} or resynthesis from audio-visual representations~\cite{Hsu_2023_CVPR}. Several recent systems condition on text predicted by a lip reader. They pass it to a visual text-to-speech model~\cite{Hegde_2023}, use it to guide diffusion through a speech recognizer~\cite{yemini2024lipvoicer}, or attend to it together with the video and a reference of the speaker's voice or face~\cite{Choi_2025,lee2026watchspeechtextawarevideotospeech}. Lip-to-speech generation has also been shown to outperform audio-visual enhancement at very low signal-to-noise ratios~\cite{richter2025lipdiffuserliptospeechgenerationconditional}. All of these methods generate speech for the whole clip. None of them is designed to keep the original speech that survives in a partly damaged recording, or to reason over its words to recover the missing ones.

\paragraph{Speech inpainting and editing.}
Audio-visual speech inpainting fills gaps of only up to 1.6\,s in a recording from the speaker's face, without a transcript~\cite{9413488,montesinos23_interspeech}. Text-guided inpainting and editing models regenerate a marked span of a recording from its transcript while keeping the voice~\cite{borsos22_interspeech,NEURIPS2023_2d8911db,peng-etal-2024-voicecraft,wang2025maskgct}. Zero-shot text-to-speech models clone a voice from a short prompt~\cite{chen-etal-2025-f5,casanova24_interspeech}. These models need the correct words as input and do not see the lips. They therefore cannot recover what was said, nor align the regenerated speech with the video. Our pipeline supplies the words from the lips and the context, and its generator attends to the video while it speaks them.

\paragraph{Audio-visual speech enhancement.}
Audio-visual enhancement and separation use the target speaker's lip motion and face to isolate their voice from noise and competing speakers~\cite{afouras2018conversation,Ephrat_2018,Gao_2021_CVPR,Rahimi_2022_CVPR}. Generative variants resynthesize the speech instead of filtering it. They use a neural vocoder~\cite{10096390} or a diffusion model~\cite{10446625}, and the vocoder-based method extends enhancement to low signal-to-noise ratios. Audio-only enhancers~\cite{schroter23b_interspeech} and restoration models that also take a transcript~\cite{10248089} cannot use the video. In our experiments, a prompt-based separation model~\cite{shi2025sam} leaves heavily corrupted speech essentially unchanged. Enhancement methods are designed to recover a voice that is still present under the interference. Our generator also starts from the damaged track, but it is additionally conditioned on a transcript, which carries the content where the audible evidence disappears.

\paragraph{Joint audio-video generation.}
Open foundation models now generate video and audio jointly in one diffusion transformer with separate video and audio streams~\cite{hacohen2026ltx}. Built on the same backbone, Just-Dub-It~\cite{Chen_2026} regenerates both the lips and the speech to dub a video into another language. Our task is complementary: the video is fixed and attended to, and only the speech is generated to match it.

\section{Method}
\label{sec:method}

Given a video of a speaker, $V$, and its corrupted audio track, $\tilde{a}$, the goal is to recover both what was said and the speech itself. We write the corruption as
\begin{equation}
  \tilde{a} \;=\; (1-\mathbf{1}_{\Omega})\odot a \;+\; \mathbf{1}_{\Omega}\odot\left(\gamma\, a + n\right),
  \label{eq:corruption}
\end{equation}
where $a$ is the clean speech, $\Omega$ is the set of corrupted time intervals with indicator $\mathbf{1}_{\Omega}$, and $n$ is an interfering signal. Silencing ($\gamma = 0$, $n = 0$) removes the speech inside $\Omega$. Noise ($\gamma = 1$) adds interference on top of it, such as babble, a competing talker or environmental sound, at an unknown signal-to-noise ratio. $\Omega$ may span a single word, several words or the whole clip. The output is a transcript $\hat{y}$ and a speech signal $\hat{a}$ that is synchronized with $V$ and keeps the speaker's voice. The videos are in the wild, so head pose, lighting and how clearly the lips can be seen all vary.

\begin{figure}[t]
  \centering
  \includegraphics[width=\linewidth]{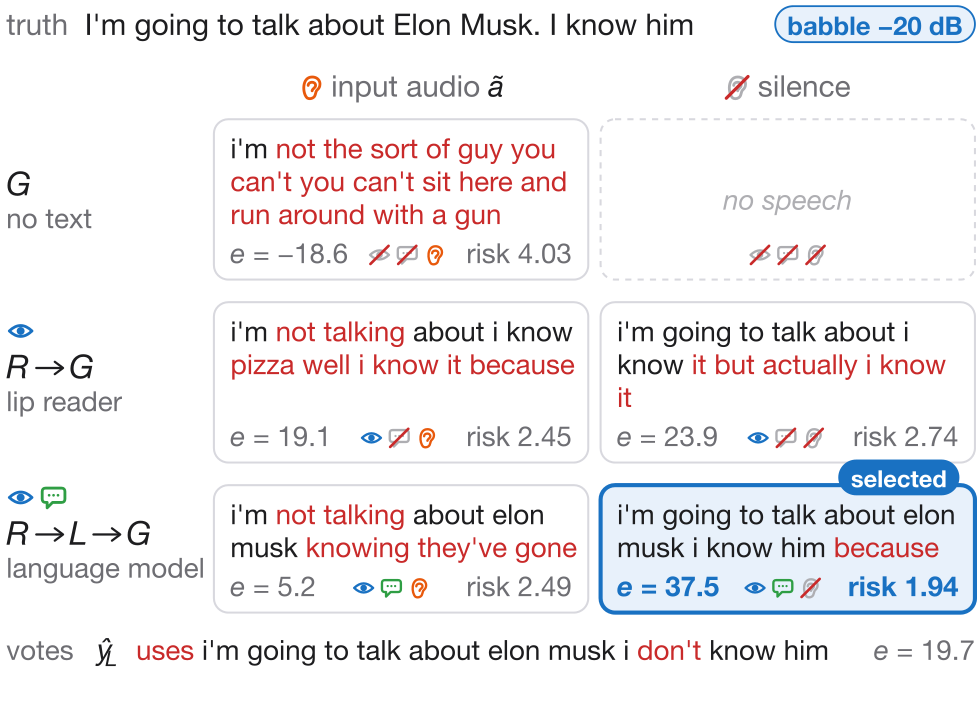}
  \caption{\textbf{Test-time selection} on a test clip with babble at $-20$\,dB. Each card is one configuration (rows: text given to $G$; columns: starting audio), with its transcript (red: wrong words), evidence $e$ (\cref{eq:evidence}) and risk (\cref{eq:mbr}). $\hat{y}_L$ only votes. In this case, every configuration that keeps the noisy audio fails, and the selection switches it off.}
  \label{fig:selection}
\end{figure}

\subsection{Overview}
\label{sec:method:overview}

 RML combines visual, linguistic, and acoustic information via three complementary models (\cref{fig:pipeline}). A lip reader $R$ (visual modality) produces a list of $N$-best transcript hypotheses $\mathcal{H}=(h_1,\ldots,h_N)$. A language model $L$ (context modality) combines these hypotheses with a Whisper transcript $c$ of $\tilde{a}$, in which silenced intervals, located by a simple energy threshold, are marked as gaps. Finally, an audio-visual generator $G$ (audio-visual modality) reconstructs the speech conditioned on $V$, $\tilde{a}$, and the recovered text. The models communicate through plain text, making each replaceable; in particular, $R$ can be replaced by any visual speech recognizer without retraining $L$ or $G$. We first describe each component of RML, and then introduce an adaptive selection process that determines how to use them based on the reliability of the available audio and visual evidence.

\subsection{RML Framework}
\label{sec:method:framework}

\noindent\textbf{Lip Reader as the Visual modality.} When the audio is missing or corrupted, a lip reader provides complementary visual evidence of what was said. We use Auto-AVSR~\citep{ma2023auto} for English and the CMU-MOSEAS readers of \citet{Ma2022} for Spanish and French; the choice of reader is not critical, as swapping in USR~2.0~\citep{haliassos2026pay} without retraining any other component keeps the same gains (\cref{sec:ablation_reader}). We use a pretrained visual speech recognizer that takes the mouth region of $V$ and returns the top $N$ transcripts using beam search. We process the full clip rather than only $\Omega$, allowing the reader to use surrounding visual clues and its hypotheses to be aligned with the audible transcript downstream. We retain the $N$-best list to provide alternative visual hypotheses to the language model, increasing the chance that the correct words are among the candidates.

\noindent\textbf{Contextual Reasoning (Language Model).} The language model (Gemma-4 12B~\cite{team2026gemma}) combines the visual hypotheses with the audible context to recover the transcript. We obtain the context $c$ by transcribing $\tilde{a}$ with Whisper large-v3~\citep{radford2023robust}, marking detected silent intervals with placeholders and providing their durations; for noisy inputs, the transcript may instead contain incorrect or missing words. The model receives the audible context $c$, a task tag (\texttt{<inpaint>} if a silent gap is detected, \texttt{<denoise>} otherwise), and a list of candidate transcripts. For silenced inputs, the candidates are the lip reader's $N$-best hypotheses $\mathcal{H}$. For noisy inputs, where the audio still carries information, we interleave $\mathcal{H}$ with the $N$-best hypotheses of the audio-visual Auto-AVSR model~\citep{ma2023auto} (or any AV recognizer) and add Whisper's transcript of the noisy audio, removing duplicates. The model first chooses the most likely candidate and then uses the surrounding context to correct it and produce the final transcript.
We fine-tune the model on silenced and noisy clips with known word timings using the best candidate and reference transcript as supervision. To avoid dependence on a particular lip reader, training hypotheses are sampled from readers of different architectures and quality, including degraded-video variants, while the number of hypotheses is sampled uniformly between one and $N$; we use $N{=}10$ throughout.

\noindent\textbf{AV modality and Speech Reconstruction.}
While any audio-visual generative model can be used, we use LTX-2.5~\cite{hacohen2026ltx} because it is open-source and publicly available. LTX-2.5 is a diffusion transformer that jointly models video and audio latents with bidirectional attention and text conditioning.
We adapt the generator using dual canvas low-rank adapters~\cite{ben2026avcontrol} and use it only to generate speech: the video latents of $V$ provide visual clues, while the audio stream contains the input track as clean context followed by the target audio to be generated. A task tag, \texttt{<inpaint>} or \texttt{<denoise>}, optionally followed by a transcript, provides text conditioning. Together, these signals allow the generator to align speech with lip motion, preserve the speaker's voice and audible speech, and reconstruct content that is silent or masked. The generated speech is synchronized with $V$. Further training and implementation details are provided in the Appendix~\cref{sec:supp:implementation}.

\subsection{Adaptive Component Selection}
\label{sec:method:selection}

The usefulness of each model depends on the input: audible speech favors the audio, missing words benefit from context, heavy noise favors visual evidence, and invisible lips make the lip reader unreliable. Rather than selecting these cases manually, we run several configurations with different components and audio conditions, and select the output best supported by the available evidence. 
Specifically, each configuration $\theta$ is a choice of the two inputs given to the generator $G$: the text and the audio track. The text is either omitted, the lip reader's top hypothesis $h_1$, or the language model's transcript $\hat{y}_L$. The audio track is either the corrupted input $\tilde{a}$, so that $G$ can reuse the audible speech, or silence, so that $G$ ignores it. We name the configurations by the components that produce the text: $G$ (no text), $R\!\rightarrow\!G$ ($h_1$), and $R\!\rightarrow\!L\!\rightarrow\!G$ ($\hat{y}_L$), each with or without $\tilde{a}$. Crossing the two choices gives six configurations (Fig.~\ref{fig:selection}). When the whole clip is silenced, $\tilde{a}$ is itself silent, so only the three text options remain. Each configuration produces a candidate speech signal $\hat{a}_\theta$ and its transcript $y_\theta$.

\noindent\textbf{Evidence-Based Scoring.} To select the best configuration, we need to determine which candidate is most consistent with the available evidence. The challenge in test-time is that we do not have any ground-truth to compare with or choose based on it. We therefore define an evidence score ($e(y)$) for each candidate transcript ($y$), which measures its support from the actual audio and lip motion while removing the recognizers' prior preference for sentences that are likely on their own. The audio and visual evidence scores, ($e_A(y)$) and ($e_R(y)$), are computed separately and then combined:
\begin{align}
  e_A(y) &= \log p_A(y \mid \tilde{a}) - \log p_A(y \mid \varnothing_a), \nonumber \\
  e_R(y) &= \log p_R(y \mid V) - \log p_R(y \mid \bar{V}), \label{eq:evidence} \\
  e(y) &= e_A(y) + e_R(y), \nonumber
\end{align}
where $\varnothing_a$ is silence and $\bar{V}$ repeats the first mouth frame, providing baselines without acoustic or visual evidence, respectively.
Subtracting these baseline scores isolates the contribution of the actual audio and lip motion to ($e_A(y))$ and ($e_R(y)$), and their sum ($e(y)$) gives the overall evidence for ($y$). Thus, when a modality is uninformative, such as heavily corrupted audio or invisible lips, its evidence score is close to zero and has little influence on the final selection.

\noindent\textbf{Final Selection.}
We select the output using minimum Bayes risk~\cite{goel2002minimum,kumar2004minimum,fiscus1997post}. The voters are the configuration transcripts and $\hat{y}_L$, except the output generated from $h_1$ without the audio ($R\rightarrow G$), which $\hat{y}_L$ already accounts for since $L$ reads the whole $N$-best list:
\begin{equation}
  \theta^\star = \arg\min_{\theta \in \Theta} \sum_{y' \in \mathcal{Y}} e^{\frac{e(y')}{T}} \, d(y_\theta, y'),
  \label{eq:mbr}
\end{equation}
where $\mathcal{Y}$ is the set of voters, $d$ is the bidirectional word error rate (WER), and $T$ is set on the validation set. $\hat{y}_L$ votes but cannot be selected. The chosen configuration provides both $\hat{a}_{\theta^\star}$ and $\hat{y} = y_{\theta^\star}$.

The selection adapts to the reliability of the available audio and video. When the audio is heavily corrupted, configurations that ignore it are more likely to be selected (\cref{fig:selection}). When the audio is clean, configurations that use it are preferred. The evidence scores also reduce the effect of predictions that are not supported by the audio or lip motion, allowing the system to choose the most reliable information without using fixed rules.
\section{Experiments}
\label{sec:experiments}

We first describe our experimental setup, and then evaluate our method and the impact of each of its components.
\begin{table}[t]
  \centering
  \small
  \setlength{\tabcolsep}{3.5pt}
    \caption{\textbf{Test sets.} Each 4.8-s window produces multiple silenced examples by varying the missing-span length $k$, and five noisy examples using different noise conditions. Silenced examples that fail ASR verification on the clean audio are discarded.}
  \begin{tabular}{@{}lrrcrr@{}}
    \toprule
    Set & Videos & Windows & $k$ & Silenced & Noisy \\
    \midrule
    TalkVid (en)     & 23 & 114 & 1--10   & 875   & 570 \\
    VoxMM (en)       & 28 & 159 & 1--10   & 1,217 & 795 \\
    Hard (en)        & 23 & 44  & 1, 3, 5 & 116   & 220 \\
    TalkVid (es, fr) & 76 & 431  & 1--4    & 401   & --  \\
    \bottomrule
  \end{tabular}
  
  \label{tab:testsets}
\end{table}

\noindent\textbf{Data.} We train on a small subset of TalkVid~\cite{chen2026talkvid}: 6{,}509 of 4.8\,s windows (English, Spanish and French) for the generator and 6{,}792 English windows for the language model, totaling fewer than 1{,}430 complete videos and 12 hours of speech. We test on unseen TalkVid videos, out-of-domain VoxMM~\cite{kwak2024voxmm,kwak2026lrs}, Spanish and French videos, and a challenging set with side views, occlusions, animated characters, and generated speakers (\cref{tab:testsets}).
We corrupt each window by either removing 1--10 consecutive words or masking the full audio with babble, sound effects, or another speaker. Hyperparameters are tuned on 99 validation videos and fixed across test sets; Appendix \cref{sec:supp:setup} provides full details.

\noindent\textbf{Metrics.} We evaluate outputs based on their audible content. We transcribe generated audio with Whisper large-v3~\cite{radford2023robust} and compare it with the reference. For silenced speech, we report \emph{exact span recovery} and \emph{word accuracy}, scoring only examples that Whisper correctly transcribes before corruption. For noisy speech, we report window-level WER. Statistical significance is determined using 95\% paired bootstrap intervals over source videos. The intervals are given in Appendix~\cref{sec:supp:significance}, a check with a second recognizer in Appendix~\cref{sec:supp:scoring_checks}, and the naturalness, speaker identity and lip synchronization of the generated speech in Appendix~\cref{sec:supp:quality}.

\noindent\textbf{Baselines.} 
To our knowledge, no existing method combines lip reading, linguistic context and the surviving audio to restore both silenced and noisy speech, adapting to whichever source is reliable for each input. We therefore evaluate word-recovery accuracy on both tasks against the closest baselines. For silenced speech, we compare with lip/AV recognizers (Auto-AVSR~\cite{ma2023auto}, USR\,2.0~\cite{haliassos2026pay}, Llama-AVSR~\cite{10889251}), speech editors given the missing words and location (F5-TTS~\cite{chen-etal-2025-f5}, VoiceCraft~\cite{peng-etal-2024-voicecraft}, AlignDiT~\cite{Choi_2025}), and LipVoicer~\cite{yemini2024lipvoicer}. For noisy speech, we compare with audio restoration (DeepFilterNet3~\cite{schroter23b_interspeech}, Sidon~\cite{nakata2026sidon}) and audio-visual separation (VisualVoice~\cite{Gao_2021_CVPR}, SAM Audio~\cite{shi2025sam}).

\subsection{Recovering Silenced Speech}
\label{sec:exp:silenced}

\begin{table}[t]
  \centering
  \footnotesize
  \setlength{\tabcolsep}{3.5pt}
    \caption{\textbf{Recovering silenced speech.} Exact span recovery and word accuracy (Acc.), in \%. \emph{Text} is the transcript a method receives, from the lip reader ($R$) or our language model ($L$); \emph{Span} marks methods that are told where the span lies.}
  \begin{tabular}{@{}llcrrrr@{}}
    \toprule
    & & & \multicolumn{3}{c}{Exact recovery} & Acc. \\
    \cmidrule(lr){4-6} \cmidrule(l){7-7}
    Method & Text & Span & TalkVid & VoxMM & Pooled & Pooled \\
    \midrule
    \multicolumn{7}{@{}l}{\emph{Text output}} \\
    Auto-AVSR~\cite{ma2023auto}                 & --  & --         & 51.0 & 16.1 & 30.7 & 59.7 \\
    USR\,2.0~\cite{haliassos2026pay}             & --  & --         & 53.5 & 23.0 & 35.8 & 64.9 \\
    Llama-AVSR~\cite{10889251}                  & --  & --         & 47.7 & 18.0 & 30.4 & 58.5 \\
    \midrule
    \multicolumn{7}{@{}l}{\emph{Speech output}} \\
    LipVoicer~\cite{yemini2024lipvoicer}        & own & --         & 43.4 & 17.3 & 28.3 & 58.6 \\
    VoiceCraft~\cite{peng-etal-2024-voicecraft} & $L$ & \checkmark & 51.8 & 18.3 & 32.3 & 59.8 \\
    F5-TTS~\cite{chen-etal-2025-f5}             & $L$ & \checkmark & 57.4 & 20.2 & 35.8 & 62.2 \\
    AlignDiT~\cite{Choi_2025}                   & $L$ & \checkmark & 56.3 & 20.9 & 35.7 & 62.8 \\
    Ours (Auto-AVSR as $R$)                                       & $L$/$R$ & --     & \underline{60.3} & \underline{24.2} & \underline{39.3} & \underline{65.7} \\
    Ours (USR\,2.0 as $R$) & $L/R$ & -- & \textbf{64.0} & \textbf{29.3} & \textbf{43.8} & \textbf{69.2} \\
    \bottomrule
  \end{tabular}

  \label{tab:silenced}
\end{table}

\Cref{tab:silenced} shows that our method achieves the best results across all word-accuracy and span-recovery metrics, whether using Auto-AVSR or the stronger USR\,2.0 as the lip reader. With Auto-AVSR, it exactly recovers $39.3\%$ of silenced spans, compared with $30.7\%$ for Auto-AVSR alone, already on par with or above the stronger USR\,2.0 itself, and with USR\,2.0 this rises to $43.8\%$, 8 points above USR\,2.0 alone. Notably, speech editors such as F5-TTS and AlignDiT are given both our language model's transcript and the span location, whereas our generator must infer the span itself, yet still performs better. These gains show that our method improves over the lip readers and benefits from jointly using visible lips and surrounding audible speech.

\subsection{Restoring Noisy and Overlapping Speech}
\label{sec:exp:noisy}

\Cref{tab:noisy} reports WER over the full 4.8\,s speech windows, averaged across the different noise conditions; results broken down by individual test set are provided in Appendix~\cref{sec:supp:noisy_sets}. On average, our pipeline reduces WER from $69.0\%$ for the noisy input to $25.5\%$, outperforming all baselines. Audio-only methods struggle under strong noise, while video-only methods discard useful speech. AV recognizers such as USR\,2.0 (AV) benefit when the audio remains less noisy and mostly informative: at $-5$\,dB babble, USR\,2.0 (AV) comes closest to ours because the relatively mild noise still allows the audio to reveal much of the speech. As corruption increases, this advantage disappears; at $-10$\,dB babble, our method substantially outperforms all AV recognizers. Under a competing talker, VisualVoice, a model built specifically for speaker separation, comes level with ours.

These results show the benefit of weighing the more reliable signals more heavily, which motivates the adaptive selection described next. \Cref{sec:exp:ablations} further shows how each signal contributes to the final performance.

\begin{table}[t]
  \centering
  \small
  \setlength{\tabcolsep}{4.0pt}
    \caption{\textbf{Restoring noisy speech.} WER (\%, lower is better) of whole 4.8\,s windows, pooled over TalkVid and VoxMM. \emph{Other talker}: a competing speaker at $-5$\,dB. Text outputs are scored directly. Speech outputs are transcribed first. V: lips only; AV: lips and noisy audio. $^\dagger$Only windows with successful preprocessing.}
  \begin{tabular}{@{}lrrrrrr@{}}
    \toprule
    & \multicolumn{2}{c}{Babble} & \multicolumn{2}{c}{Sound effects} & Other & \\
    \cmidrule(lr){2-3} \cmidrule(lr){4-5} \cmidrule(lr){6-6}
    Method & $-10$ & $-5$ & seen & unseen & talker & Avg \\
    \midrule
    Noisy input                                          & 91.8 & 52.9 & 47.7 & 48.0 & 104.8 & 69.0 \\
    \midrule
    \multicolumn{7}{@{}l}{\emph{Text output}} \\
    Auto-AVSR~\cite{ma2023auto} (V)                      & 49.4 & 49.4 & 49.4 & 49.4 & 49.4  & 49.4 \\
    Auto-AVSR~\cite{ma2023auto} (AV)                     & 82.1 & 47.7 & 68.7 & 61.0 & 93.6  & 70.6 \\
    USR\,2.0~\cite{haliassos2026pay} (V)                  & 47.4 & 47.4 & 47.4 & 47.4 & 47.4  & 47.4 \\
    USR\,2.0~\cite{haliassos2026pay} (AV)                 & 62.4 & 28.8 & 35.4 & 29.0 & 104.3 & 52.0 \\
        Llama-AVSR~\cite{10889251} (V)                       & 58.8 & 58.8 & 58.8 & 58.8 & 58.8  & 58.8 \\
    Llama-AVSR~\cite{10889251} (AV)                      & 57.8 & 34.3 & 30.7 & 26.0 & 90.4  & 47.9 \\
    \midrule
    \multicolumn{7}{@{}l}{\emph{Speech output}} \\
    DeepFilterNet3~\cite{schroter23b_interspeech}        & 92.5 & 70.6 & 51.6 & 43.3 & 102.8 & 72.2 \\
    Sidon~\cite{nakata2026sidon}                         & 99.4 & 93.0 & 61.5 & 64.8 & 103.1 & 84.4 \\
    VisualVoice~\cite{Gao_2021_CVPR}$^\dagger$           & 80.5 & 41.8 & 28.7 & 29.1 & \textbf{20.7} & 40.1 \\
    LipVoicer~\cite{yemini2024lipvoicer}                 & 52.9 & 52.9 & 52.9 & 52.9 & 52.9  & 52.9 \\
    AlignDiT~\cite{Choi_2025} ($R$)$^\dagger$            & 50.5 & 50.5 & 50.5 & 50.5 & 50.5  & 50.5 \\
    Ours (Auto-AVSR as $R$)  & \underline{44.0} & \underline{28.1} & \underline{22.0} & \underline{19.9} & 22.1 & \underline{27.2} \\
    Ours (USR\,2.0 as $R$)       & \textbf{40.1} & \textbf{26.6} & \textbf{20.6} & \textbf{19.0} & \underline{21.1} & \textbf{25.5} \\
    \bottomrule
  \end{tabular}

  \label{tab:noisy}
\end{table}

\subsection{Adapting to Audio Corruption}
\label{sec:exp:adapt}

Different types and levels of corruption make different signals more reliable, but at test time we do not know which configuration is best for a given input. Our method therefore considers all six configurations and uses their evidence scores to select the most reliable output. To test this, we use the same test windows and vary both the corruption and its duration: babble at $-5$, $-10$, or $-20$\,dB, or silence, affects one word, four words, or the whole clip. We also include clean clips. This gives 1,421 TalkVid and 1,998 VoxMM samples, evaluated using WER over the full window. 

\begin{figure}[t]
  \centering
  \includegraphics[width=\linewidth]{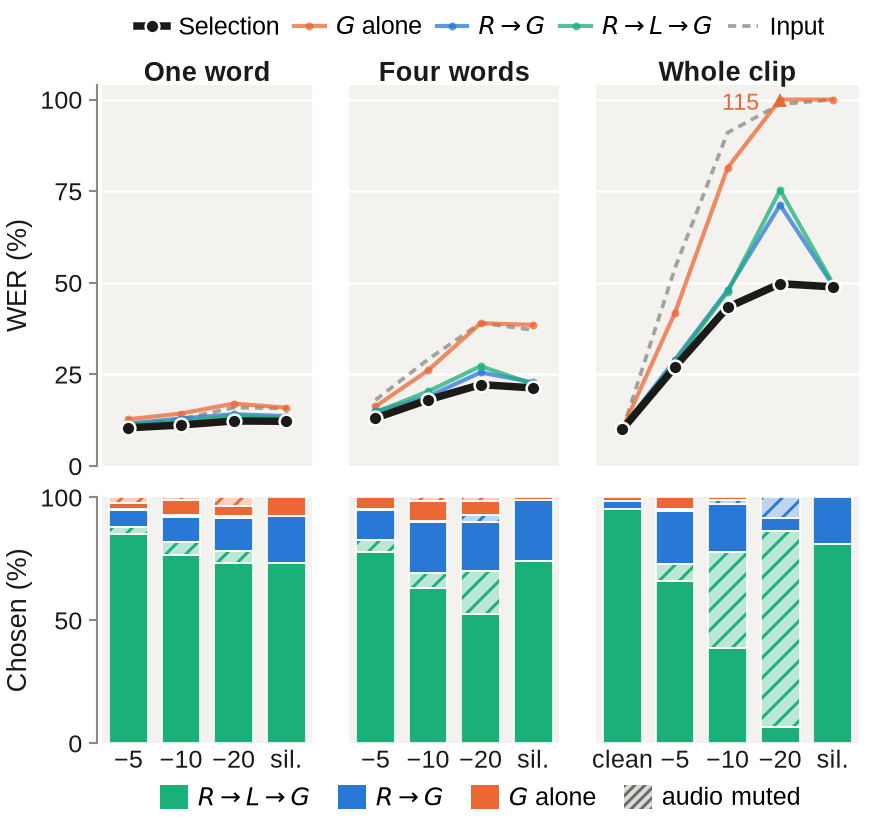}
      \caption{\textbf{Adapting to the input.} WER under different corruption levels and durations. Top: the black line shows our adaptive selection, which matches or improves on the best fixed configuration, while no fixed configuration performs best across all scenarios. Bottom: different configurations are selected as the corruption changes; stripes indicate that the input audio is replaced with silence.}
  \label{fig:adapt}
\end{figure}

\cref{fig:adapt} (Bottom) shows how our method changes its selection as the audio becomes less reliable. For clean clips, it almost always keeps the audio. As the corruption becomes stronger and covers more of the clip, it increasingly selects configurations that replace the corrupted audio with silence. For example, when the whole clip is corrupted by $-20$\,dB babble (right diagram), it removes the audio for 88\% of the clips, and the voice is then inferred from the speaker's appearance rather than taken from the recording (Appendix~\cref{sec:supp:quality}). Importantly, the method is not given the noise level or corrupted region; it makes this decision from the evidence scores. The audio evidence drops from $\sim80$ for clean clips to $\sim3$ at $-20$\,dB, giving more weight to the transcript and visual evidence.

\cref{fig:adapt} (Top) shows the benefit of this selection. The black curve shows our adaptive selection, which matches or improves on the best fixed configuration across the different corruption conditions. No single fixed configuration works best across all corruption levels. The generator alone works well when the noise is mild, but becomes unreliable under severe corruption, reaching over 100\% WER when the whole clip is corrupted at $-20$\,dB. Using a transcript helps, but configurations that keep the corrupted audio still suffer from the noise. By choosing whether to keep or remove the audio based on its reliability, our method reaches $49.8\%$ WER in this difficult case. Across all conditions, adaptive selection achieves $23.1\%$ WER, compared with $26.4\%$ for always using $R \rightarrow G$, $26.8\%$ for $R \rightarrow L \rightarrow G$, $40.8\%$ for the generator alone, and $49.1\%$ for the lip reader.

\subsection{Hard-to-Read Lips}
\label{sec:exp:hard}

Here we test whether our method remains robust when lip reading is unreliable by detecting weak visual evidence and shifting toward more reliable sources, such as language and audio. We collect $44$ English windows from $23$ scenes (e.g., movies, animation and generated videos) in which the mouth is \emph{hard or impossible} to read because of side views, occlusions or animated characters. On these windows we silence spans of $k=1$, 3 and 5 words (116 verified cuts) and apply the noise conditions of \cref{tab:noisy} (220 noisy samples), and compare with baselines on both tasks (\cref{tab:hard}).

Our method scores highest on both tasks, recovering three times more silenced spans than its own lip reader. When the lips are unreliable, it relies more on the other components: the language model predicts the missing words from context, and the generator speaks them in the speaker's voice. This also explains why speech editors given our language model transcript perform closest to our method on silenced speech. Importantly, our method is not told when lip reading fails; the language model and the surviving audio compensate for what the lips cannot show. Lip-reading baselines use only the lips ($V$) for silenced speech and both lips and audio ($AV$) for noisy speech, but still perform much worse. Some baselines (marked $^\ast$) require a detected face and fail when the mouth is never visible, so we score them only on the windows they can process.

\begin{table}[t]
  \centering
  \small
  \setlength{\tabcolsep}{10.0pt}
  \newcommand*{\fp}{\rlap{$^\ast$}}
    \caption{\textbf{Hard-to-read lips.} Exact span recovery (\%) on the 116 silenced cuts and WER (\%) over the 220 noisy samples. Methods marked $^\ast$ are scored only where their face preprocessing succeeds (matched comparisons in Appendix~\cref{sec:supp:significance}).}
  \begin{tabular}{@{}lrr@{\hspace{6pt}}}
    \toprule
    Method & Silenced $\uparrow$ & Noisy $\downarrow$ \\
    \midrule
    Damaged input                                  & --        & 54.1 \\
    Auto-AVSR~\cite{ma2023auto} ($V$ / $AV$)              & 8.6       & 72.5\fp \\
    USR\,2.0~\cite{haliassos2026pay}                & 12.0\fp   & 56.3\fp \\
    Llama-AVSR~\cite{10889251}                     & 13.3\fp   & 60.3\fp \\
    SAM Audio~\cite{shi2025sam}, text prompt       & --        & 53.7 \\
    VisualVoice~\cite{Gao_2021_CVPR}               & --        & 41.6\fp \\
    LipVoicer~\cite{yemini2024lipvoicer}           & 10.8\fp   & 92.8\fp \\
    F5-TTS~\cite{chen-etal-2025-f5} ($L$)          & 23.3      & -- \\
    AlignDiT~\cite{Choi_2025} ($L$ / $R$)          & 23.6\fp   & 97.0\fp \\
    \midrule
    Ours (Auto-AVSR as $R$)                        & \textbf{25.9} & \textbf{29.6} \\
    \bottomrule
  \end{tabular}

  \label{tab:hard}
\end{table}

\subsection{Generalization}
\label{sec:ablation_reader}

Since the components communicate via text, $R$ can be replaced with no retraining. With USR\,2.0 as $R$ under the same selection (\cref{tab:silenced,tab:noisy}), our method improves on both tasks: a stronger lip reader directly yields a stronger pipeline, while $L$ and $G$ provide further gains. Interestingly, even with Auto-AVSR as $R$, our method remains the second-best overall, behind only our version with USR\,2.0. This shows that the gains come from the framework itself and remain strong across different lip readers. The same holds across languages (\cref{tab:ablations}b): with a Spanish or French lip reader as $R$, and $L$ and $G$ unchanged, the pipeline improves exact recovery over the lip reader by a third.

\subsection{Ablations}
\label{sec:exp:ablations}

\paragraph{Components.} \cref{tab:ablations} ablates the components of our pipeline to demonstrate the contribution of each model. First, giving the lip reader's $N$-best hypotheses to the language model, together with the audible context, gives a much better transcript than the lip reader alone ($R \rightarrow L$ against $R$). Second, the generator $G$ is more than a renderer. If $G$ simply spoke the input text like a text-to-speech model, the lips it sees would not matter. Instead, with the same transcript but mismatched lips, it performs significantly worse (Appendix~\cref{sec:supp:checks}).
Third, the language model matters most for long gaps: without it ($R \rightarrow G$) the pipeline loses spans at every length, and for seven to ten missing words the full pipeline recovers 42\% more spans than the lip reader.

No fixed configuration is best everywhere. The language model helps on silenced speech, but on noisy speech the audio still carries most of the words, and $R \rightarrow G$ does better without it. Our selection does not know which case it is in, yet it gives the lowest error on noisy speech and is close to the best on silenced speech. The only exception is the silenced hard set, where the difference is not significant. Here the lips cannot be read, making the visual evidence used by the selection unreliable and sometimes causing it to choose an output that follows the lip reader's wrong prediction.

\paragraph{Lip features for the language model.} One may expect the language model to read better if it saw the lips directly rather than through a lip reader's hypotheses. We test this with a small head that reads AV-HuBERT~\cite{shi2022learning} mouth features and passes its viseme posteriors to $L$ as soft input tokens, trained jointly with $L$. The model does rely on the head, since features from another clip lower exact recovery by 22.8 points on the validation set, yet the head brings no gain on any test set and loses spans where the lip features are least reliable (\cref{tab:ablations}c). The ranked hypotheses thus carry the visual information $L$ needs.

\paragraph{Selection rule.} To check that every term of the selection rule (\cref{eq:mbr}) is needed, we remove them one at a time. Each helps in at least one setting, and removing any term gives no significant improvement in any setting (Appendix~\cref{sec:supp:selection}).

\paragraph{Remaining errors.} Given the true transcript, $G$ recovers 96.6\% of the silenced spans on TalkVid and VoxMM, more than any speech editor given the same text. Under noise, it roughly halves the WER, bringing it close to that of the clean recordings; the hard set shows the same pattern. This suggests that most remaining errors come from the transcript and not the generator, making the lip reader and language model the main components to improve.

\begin{table}[t]
  \centering
  \small
  \begin{tabular*}{\linewidth}{@{\extracolsep{\fill}}lrrrr@{}}
    \toprule
    \multicolumn{5}{@{}l}{\emph{(a) Components}} \\
    & \multicolumn{2}{c}{Silenced $\uparrow$} & \multicolumn{2}{c}{Noisy $\downarrow$} \\
    \cmidrule(lr){2-3} \cmidrule(l){4-5}
    & Pooled & Hard & Pooled & Hard \\
    \midrule
    $R$ (text)                       & 30.7 & 8.6  & 49.4 & 91.4 \\
    $R \rightarrow L$ (text)         & 36.2 & 22.4 & 42.8 & 74.5 \\
    $G$ without text                 & --   & --   & 45.2 & 37.6 \\
    $R \rightarrow G$                & 33.1 & 14.7 & \underline{29.0} & \underline{30.6} \\
    $R \rightarrow L \rightarrow G$  & \underline{38.6} & \textbf{27.6} & 29.5 & 35.2 \\
    Selection (ours)                 & \textbf{39.3} & \underline{25.9} & \textbf{27.2} & \textbf{29.6} \\
    \bottomrule
  \end{tabular*}
  
  \vspace{0.3em}
  \begin{tabular*}{\linewidth}{@{\extracolsep{\fill}}lrrr@{}}
    \toprule
    \multicolumn{4}{@{}l}{\emph{(b) Spanish and French}, exact span recovery (\%) $\uparrow$} \\
    & Spanish & French & Pooled \\
    \midrule
    \multicolumn{4}{@{}l}{\emph{$R$ = Ma~\etal~\cite{Ma2022}}} \\
    $R$ (text)                       & 34.0 & 17.5 & 30.7 \\
    $R \rightarrow L$ (text)         & \underline{36.8} & \underline{21.2} & \underline{33.7} \\
    $R \rightarrow L \rightarrow G$  & \textbf{43.9} & \textbf{28.8} & \textbf{40.9} \\
    \bottomrule
  \end{tabular*}
  
  \vspace{0.3em}
  \begin{tabular*}{\linewidth}{@{\extracolsep{\fill}}lrrrr@{}}
    \toprule
    \multicolumn{5}{@{}l}{\emph{(c) Input of $L$}, exact span recovery (\%) $\uparrow$} \\
    & TalkVid & VoxMM & Hard & es, fr \\
    \midrule
    $N$-best list (ours)              & 60.5 & 22.9 & 27.6 & 40.9 \\
    $N$-best list + lip features      & 60.6 & 22.9 & 25.9 & 38.4 \\
    \bottomrule
  \end{tabular*}
\caption{\textbf{Ablations.} (a)~Each configuration of our pipeline with Auto-AVSR as $R$: exact span recovery (\%) on the silenced cuts and WER (\%) on the noisy windows, pooled over TalkVid and VoxMM, and on the hard set. (b)~Spanish and French: only $R$ is replaced, by the lip reader of Ma~\etal~\cite{Ma2022}; $L$ and $G$ are unchanged. (c)~$R \rightarrow L \rightarrow G$ with and without a lip-feature head in the language model.}
  \label{tab:ablations}
\end{table}

\section{Conclusions}

We introduced Read My Lips, a method for recovering speech by bringing together lip reading, language understanding, and audio-visual generation. This work points to an intriguing opportunity made possible by recent advances in language and audio-visual models: we can now combine their complementary abilities to recover information that each model alone may miss. Words that are unclear from the lips may become clear from the conversation, and speech buried in noise may still be revealed by the speaker’s mouth movements. This is similar to human perception, focusing on clearest sensory cues based on the situation. Together, these clues can make an otherwise unintelligible recording understandable again. 

Making this collaboration work requires knowing when to trust each source. More information is not always better: severely corrupted audio can mislead the model, while ignoring it can improve recovery. Similarly, a language model can produce a convincing sentence without knowing whether it was actually spoken, and lip reading can become unreliable when the mouth is unclear or occluded. The challenge is that at inference time, it is not known which of these sources is reliable for a given input. Their reliability must therefore be judged from the available evidence. This creates an opportunity to combine the language model's ability to reason about speech with the evidence in the audio and lip motion, relying on each source when it is informative and less when it is not.

What remains unsolved is also encouraging. Once the words are correct, the generator can reconstruct the speech with high accuracy. As speech generation becomes increasingly capable, recovering the correct words becomes the central challenge. The emergence of strong language and audio-visual models makes this a timely direction to explore: how far can we go in recovering what was said by combining what can be heard, seen, and inferred?

{
    \small
    \bibliographystyle{ieeenat_fullname}
    \bibliography{main}
}

\clearpage
\renewcommand{\textfraction}{0.15}
\setcounter{page}{1}
\maketitlesupplementary
\appendix

\section{Setup}
\label{sec:supp:setup}

This section details the evaluation sets, corruptions, scoring, and baseline implementations. Pairwise comparisons and confidence intervals are reported in \cref{sec:supp:results}.
 
\subsection{Test sets}
\label{sec:supp:testsets}

\begin{table}[htbp]
  \centering
  \footnotesize
  \setlength{\tabcolsep}{3.2pt}
  \caption{\textbf{Evaluation sets.} A silenced cut removes $k$ consecutive words from one window. Only \emph{verified} cuts, whose words Whisper transcribes exactly from the original audio, are scored. Each window gives one noisy sample per interference condition. $^\dagger$Scenes.}
  \begin{tabular}{@{}lrrcrrr@{}}
    \toprule
    & & & & \multicolumn{2}{c}{Silenced cuts} & Noisy \\
    \cmidrule(lr){5-6}
    Set & Videos & Windows & $k$ & all & verified & samples \\
    \midrule
    TalkVid (en)      & 23  & 114 & 1--10   & 958   & 875   & 570 \\
    VoxMM (en)        & 28  & 159 & 1--10   & 1,424 & 1,217 & 795 \\
    Hard (en)         & 23$^\dagger$ & 44 & 1, 3, 5 & 119 & 116 & 220 \\
    TalkVid (es, fr)  & 76  & 431  & 1--4    & 431   & 401   & -- \\
    \midrule
    Validation (en)   & 99  & 99  & 1--10   & 373   & 360   & 495 \\
    \bottomrule
  \end{tabular}
  \label{tab:supp:testsets}
\end{table}

All test windows are 4.8\,s long and come from videos excluded from training for every component (\cref{tab:supp:testsets}). The VoxMM~\cite{kwak2024voxmm,kwak2026lrs} windows come from its test split. These conversational recordings come from a different source than TalkVid, and no VoxMM data trains the language model or generator. The Spanish and French set has 334 Spanish and 97 French windows, from 69 and 7 TalkVid videos, and each window holds a single cut.

The hard set contains English windows in which the mouth is difficult or impossible to read because the speaker is viewed from the side, the mouth is hidden by a hand or an object, or the speaker is animated or generated. Twenty-eight windows come from generated videos, and the other 16 from films and animation. As one measure of visual availability, we run a face landmark detector on every frame. It finds the mouth in every frame of 32 windows, in some frames of 5, and in no frame of the remaining 7.

All hyperparameters, including the temperature $T$ of the selection, are chosen on a validation set of English TalkVid windows, each from a different video, that shares no video with the training or test sets.

\subsection{Corruptions}
\label{sec:supp:corruptions}

\paragraph{Silenced speech.} For each window and each $k$ from 1 to 10, we seek one span of $k$ consecutive words using forced-alignment timings. We exclude spans that begin or end with a stop word or filler, whose boundaries the aligner places least reliably. At least one word must remain audible on either side, keeping the span away from window edges where words may be truncated. We set the audio to zero from 80\,ms before the span's first word to 80\,ms after its last, stopping 20\,ms short of neighboring words. We reject a position if this interval leaves part of a span word audible or if a word at the span edge lasts less than 80\,ms, indicating a faulty alignment. Among valid positions, we prefer one containing the word around which the window was cut, then the one nearest the window center. A window yields no cut for a given $k$ if no position meets these rules (\cref{tab:supp:testsets}). For TalkVid, and for Spanish and French, the transcripts and word timings come from WhisperX. For VoxMM we use the transcripts and word timings of the LRS-VoxMM release unchanged. Where a manual review corrected a reference, the corrected words take the time span of the words they replace, and cuts that cover only part of a corrected run are removed.

\paragraph{Noisy speech.} For babble, training uses DEMAND cafeteria noise~\cite{thiemann2013diverse} from microphone channels 1 to 12. For testing, we mix each full window with a variety of interference, to examine how the method behaves under different conditions. The main evaluation has five conditions. Two use DEMAND cafeteria babble from the held-out channels 13 to 16, at $-10$ and $-5$\,dB. Two use VGGSound environmental sounds~\cite{chen2020vggsound} at $-5$\,dB, from test clips of the nine classes used in training or from three held-out classes (church bell, sneezing, train horn). The fifth is a competing speaker at $-5$\,dB, drawn from another video in the same test set. \Cref{sec:supp:babble} repeats the babble evaluation with two independent sources, the DEMAND restaurant scene and a mixture of six other speakers, and our WER does not change significantly. We set the signal-to-noise ratio using the root-mean-square levels of the clean window and the noise.

\paragraph{Condition grid.} For \cref{sec:exp:adapt}, we vary both the strength and extent of corruption. Babble at $-5$, $-10$ or $-20$\,dB, or silence, affects one word, four words or the whole window, and clean windows are included. The resulting grid contains 1,421 TalkVid and 1,998 VoxMM samples, and \cref{tab:supp:grid} reports every condition. We construct the validation grid in the same way from validation windows.

\subsection{Scoring and statistics}
\label{sec:supp:scoring}

Every output is transcribed over the whole window with Whisper large-v3~\cite{radford2023robust}, and the transcript is aligned with the reference after text normalization. A silenced cut counts as \emph{exactly recovered} when every silenced word is matched in place and no word is inserted inside the span. The \emph{word accuracy} is the percentage of silenced words that are matched in place. Methods that only produce text, such as the lip readers, are scored on their text with the same alignment.

Whisper can itself misrecognize disfluencies, ambiguously written numbers, or words truncated at a window edge. To avoid counting such errors as restoration failures, we score a silenced cut only when Whisper transcribes its missing words exactly from the original, undamaged audio. These \emph{verified} cuts account for 91\% of TalkVid cuts and 85\% of VoxMM cuts (\cref{tab:supp:testsets}). For noisy speech, we report whole-window word error rate averaged over windows. It can exceed 100\% when an output inserts many words, for example by including a competing speaker.

Because cuts and windows from the same video are correlated, we compute 95\% percentile confidence intervals with a paired bootstrap over source videos, or over scenes for the hard set, using at least 2,000 resamples. We call a difference significant when its interval excludes zero.

\subsection{Baselines}
\label{sec:supp:baselines}

Every baseline runs with its official code and checkpoint on the same windows as our method and is scored by the same scorer.

\paragraph{Recognizers.} Auto-AVSR~\cite{ma2023auto}, USR\,2.0~\cite{haliassos2026pay} and Llama-AVSR~\cite{10889251} produce only text. For silenced speech they read the lips alone. For noisy speech we run each from the lips alone (V) and from the lips together with the noisy audio (AV). The visual Auto-AVSR model is also our lip reader $R$.

\paragraph{Speech editors.} F5-TTS~\cite{chen-etal-2025-f5} and VoiceCraft~\cite{peng-etal-2024-voicecraft} fill a marked audio span with supplied words in the speaker's voice, and AlignDiT~\cite{Choi_2025} also conditions on the lips. Each editor receives our language model's transcript, the same text available to our generator, as well as the silenced span's location, which our method must infer from the input. We score the output returned by each editor's official inference code. Pasting only AlignDiT's generated span back into the input changes its exact recovery from 35.7\% to 36.6\% without affecting any comparison. With the lip reader's best hypothesis in place of the language model's transcript, F5-TTS and AlignDiT recover 30.3\% and 31.4\% of spans exactly, about five points fewer than with $L$'s transcript (\cref{tab:supp:ci}a).

\paragraph{Video-to-speech.} LipVoicer~\cite{yemini2024lipvoicer} synthesizes the speech of the whole window from the video, guided by its own lip reader, and ignores the input audio. On noisy speech, AlignDiT receives the lip reader's best hypothesis and, as a reference for the voice, a clean recording of the same speaker from another window. It also ignores the noisy input, so its error does not depend on the noise condition.

\paragraph{Enhancement and separation.} DeepFilterNet3~\cite{schroter23b_interspeech} and Sidon~\cite{nakata2026sidon} process the noisy audio alone. VisualVoice~\cite{Gao_2021_CVPR} separates the target speaker using their face. SAM Audio~\cite{shi2025sam} is prompted either with the text ``speech'' or with the masked video of the target speaker.

\begin{table}[htbp]
  \centering
  \footnotesize
  \setlength{\tabcolsep}{3.2pt}
  \caption{\textbf{What each method receives.} \emph{Span}: the location of the silenced interval. \emph{Transcript}: a transcript supplied from outside the method, from our lip reader ($R$) or our language model ($L$). \emph{Voice}: a clean recording of the speaker from another window. $^\ast$A text description of the target sound, not a transcript.}
  \begin{tabular}{@{}lccccc@{}}
    \toprule
    Method & Video & Audio & Span & Transcript & Voice \\
    \midrule
    Recognizers (V)            & \checkmark & --         & --         & --   & -- \\
    Recognizers (AV)           & \checkmark & \checkmark & --         & --   & -- \\
    VoiceCraft, F5-TTS         & --         & \checkmark & \checkmark & $L$  & -- \\
    AlignDiT, silenced         & \checkmark & \checkmark & \checkmark & $L$  & -- \\
    AlignDiT, noisy            & \checkmark & --         & --         & $R$  & \checkmark \\
    LipVoicer                  & \checkmark & --         & --         & --   & -- \\
    DeepFilterNet3, Sidon      & --         & \checkmark & --         & --   & -- \\
    VisualVoice                & \checkmark & \checkmark & --         & --   & -- \\
    SAM Audio, video prompt    & \checkmark & \checkmark & --         & --   & -- \\
    SAM Audio, text prompt     & --         & \checkmark & --         & --$^\ast$ & -- \\
    \midrule
    Ours                       & \checkmark & \checkmark & --         & --   & -- \\
    \bottomrule
  \end{tabular}
  \label{tab:supp:inputs}
\end{table}

\paragraph{Inputs.} \Cref{tab:supp:inputs} specifies the inputs available to each method. Our method receives the video and the damaged audio. The speech editors additionally receive the silenced span's location and our language model's transcript. For noisy speech, AlignDiT also receives a clean recording of the speaker. These additional inputs should be considered when interpreting the comparisons.

\paragraph{Coverage.} Face-dependent methods cannot process every window. On TalkVid and VoxMM, VisualVoice covers 1,245 of the 1,365 noisy windows and AlignDiT covers 1,315. Coverage is lower on the hard set, where USR\,2.0 and Llama-AVSR process about three quarters of the cuts and windows, and VisualVoice fewer than half of the noisy windows (\cref{tab:supp:hard}). For each such baseline, we compute both scores on the cuts or windows it covers.

\FloatBarrier
\section{Implementation}
\label{sec:supp:implementation}

\subsection{Models and training}
\label{sec:supp:models}

\paragraph{Lip reader and audible context.} The lip reader $R$ applies visual Auto-AVSR~\cite{ma2023auto} to mouth crops from the whole window. Beam search returns its $N = 10$ best distinct transcripts. For Spanish and French, we use the readers of Ma~\etal~\cite{Ma2022} instead. Whisper large-v3 transcribes $\tilde{a}$ to provide audible context $c$. To mark a gap in that context, we detect silent intervals directly in $\tilde{a}$ using 5\,ms frames, a threshold 70\,dB below the track peak, and a minimum duration of 50\,ms. If several intervals qualify, we choose the longest with audio on both sides. On the test sets, this detector finds the silenced interval in 99.9\% of cuts, and both boundaries are within 10\,ms of the true boundaries for 99.2\% of TalkVid cuts and 99.6\% of VoxMM cuts. \Cref{sec:supp:nonzero_gaps} examines gaps that are not digital silence and are filled instead with room tone or comfort noise. The interval is represented in $c$ by a placeholder and its duration, as illustrated in \cref{fig:supp:prompt}.

\begin{figure}[t]
  \centering
  \newcommand{\pl}[1]{\par\hangindent=1.3em\hangafter=1 #1}
  \fbox{\parbox{0.95\linewidth}{\ttfamily\scriptsize\raggedright\setlength{\parskip}{0.6pt}
    \pl{Transcript: first thing that I'm gonna do is roll a \_\_\_ paper for this candle holder}
    \pl{Lip reading candidates (best first):}
    \pl{A. the first thing that i'm going to do is throw a lot of white paper for this}
    \pl{B. the first thing that i'm going to do is throw a lot of white paper for this cat}
    \pl{C. the first thing that i want to do is throw a lot of white paper for this}
    \pl{D. the first thing that i'm going to do is throw a lot of white paper for this category}
    \pl{E. the first thing that i'm going to do is throw a lot of white paper for this cattle}
    \pl{F. the first thing that i'm going to do is control a cattle out of white paper for this cattle}
    \pl{G. the first thing that i'm going to do is scroll on cattle out of white paper for this cattle}
    \pl{H. the first thing that i'm going to do is scroll a category of white paper for this category}
    \pl{I. the first thing that i'm going to do is scroll a cattle out of white paper for this cattle}
    \pl{J. the first thing that i'm going to do is control a cattle out of wax paper for this cattle}
    \pl{Gap duration: 0.9 s}
    \vspace{3pt}
    \pl{\textbf{Pick: F}}
    \pl{\textbf{Corrected: the first thing that i'm going to do is roll a candle out of wax paper for this}}
  }}
  \caption{\textbf{A prompt of the language model and its answer} (in bold) for a TalkVid test cut whose silenced words are ``candle out''. No hypothesis contains these words. The model names the closest one and rewrites it using the context.}
  \label{fig:supp:prompt}
\end{figure}

\paragraph{Language model.} $L$ is Gemma-4 12B~\cite{team2026gemma} (instruction-tuned), adapted with LoRA~\cite{hu2021lora} of rank 32 ($\alpha = 64$, dropout 0.1) on the query, key, value and output projections of every attention layer. It is trained on English TalkVid windows only, with an effective batch of 16 (four samples and four accumulation steps) and a cosine learning-rate schedule, in three stages. In the first stage (3,000 steps, learning rate $10^{-4}$) the model learns to name the best candidate or to write the transcript. In the second (3,000 steps, $5 \cdot 10^{-5}$) it learns the answer described in \cref{sec:method:framework}, which names the best candidate and then edits it. In the third (1,000 steps, $5 \cdot 10^{-5}$) noisy examples are added to the silenced ones. The hypotheses shown during training come from four readers: Auto-AVSR on the original video (50\% of the examples), Auto-AVSR on mouth crops downscaled from 96 to 24 pixels and scaled back up (15\%), Auto-AVSR on video in which half of the frames are replaced by the preceding frame (15\%), and AV-HuBERT~\cite{shi2022learning} (20\%). The number of hypotheses shown is drawn uniformly between 1 and $N$. At test time we decode greedily.

\paragraph{Lip-feature head.} For the ablation of \cref{tab:ablations}c, a small head reads AV-HuBERT~\cite{shi2022learning} mouth features and passes its viseme posteriors to $L$ as soft input tokens. It is trained jointly with $L$. To check that the model uses these features, we replace them at test time with the features of another clip. On the 360 verified validation cuts, the exact recovery of $R \rightarrow L \rightarrow G$ then falls from 64.4\% to 41.7\% ($-22.8$, CI $[-28.4, -17.2]$).

\paragraph{Generator.} $G$ is the 22B-parameter LTX-2.5 transformer~\cite{hacohen2026ltx}, adapted with LoRA of rank 64 ($\alpha = 64$) on the audio self-attention, the audio text cross-attention, the audio feed-forward layers and the video-to-audio attention. The video pathway is left unchanged. It is trained with AdamW~\cite{loshchilov2017decoupled} at a learning rate of $10^{-4}$ and a batch size of 2 on one H100 GPU, on English, Spanish and French TalkVid, in three stages. The first stage (5,000 steps) uses about 15,800 silenced spans of one to ten words, each given the lip reader's transcript. The second (3,000 steps) uses 13,018 samples, one silenced and one noisy version of each of 6,509 windows from about 1,350 videos. Here the noisy samples receive the task tag without a transcript. Noisy training samples are mixed with babble, the seen sound effects, colored noise or a competing speaker at a signal-to-noise ratio drawn uniformly between $-15$ and $5$\,dB, sometimes with a second source 5\,dB weaker. The third stage (2,000 steps) uses the same samples but varies the text given with the noisy ones: the lip reader's transcript for half of them, the reference transcript for a quarter and no transcript for the rest. The generator thus learns to treat the transcript as a fallible hint and to work without one. The text is dropped for 10\% of all samples, which trains the unconditional branch used for classifier-free guidance. The loss weight inside the silenced region is $\lambda = 12$, applied up to 60\,ms beyond its edges. At inference we use 30 denoising steps with a fixed seed, and text guidance of 5 with \texttt{<inpaint>} and 1 with \texttt{<denoise>}. The generator produces the whole window: the audio outside the damaged region is generated together with the restored part, not copied from the input, and every score is computed on the generated window. \Cref{sec:supp:scoring_checks} shows that this leaves the intelligibility of intact speech unchanged.

\paragraph{The generator is never told where the damage is.} At inference, $G$ receives the video, the damaged track, a transcript of the whole window and the task tag. It is given neither the position nor the length of the damage, as a mask, as timestamps or as a marker in its text, and has to find by itself where speech is missing and where it must be kept. The same holds under noise: no component is told which part of a window is corrupted or how strongly, and the selection decides from the evidence alone (\cref{sec:supp:selection}). The only place where the location of a gap is used is the audible context of the language model. There, a silent interval is marked by a placeholder and its duration, because a recognizer such as Whisper does not report where words are missing from its transcript. The same detection also sets the task tag, \texttt{<inpaint>} when a silent interval is found and \texttt{<denoise>} otherwise. This detection step is a front end that can be replaced without changing the rest of the pipeline. \Cref{sec:supp:nonzero_gaps} already replaces it with a detector that uses both the audio and the lips, with nothing else changed. A recognizer that marks where it hears no speech, or that reports a confidence for each word, could supply the placeholder directly.

\subsection{Resources and runtime}
\label{sec:supp:runtime}

All training and inference runs use one NVIDIA H100 GPU with 80\,GB of memory per job. The generator's 22B-parameter backbone stays frozen in bfloat16, and only its adapters are trained. We encode the training windows' audio and video latents once in advance. A generator training step with batch size 2 then takes about 2\,s, and its three stages take about 6 GPU-hours in total. The language model takes about 2.5\,s per step on one GPU, or about 5 GPU-hours across its three stages. At inference, one 30-step generation of a 4.8\,s speech window takes about 11\,s. A fixed configuration requires one pass. Selection requires one pass for each configuration, six for a damaged window and three when no speech is audible. Each distinct resulting transcript is then scored by Whisper and the lip reader with both the real and uninformative inputs (\cref{sec:supp:selection}).

\FloatBarrier
\section{Selection}
\label{sec:supp:selection}

We detail the test-time selection introduced in \cref{sec:method:selection}, examine its choices, ablate its terms, and report the condition grid of \cref{sec:exp:adapt} in full.

\paragraph{Evidence.} Both terms in \cref{eq:evidence} are teacher-forced log-likelihoods of each candidate transcript, summed over tokens. The audio term $e_A$ uses Whisper large-v3 to score the transcript conditioned first on the damaged input $\tilde{a}$ and then on a silent track of equal length. The visual term $e_R$ uses visual Auto-AVSR's weighted attention and CTC log-likelihoods ($0.9$ and $0.1$, its decoder weights), conditioned first on the window's mouth crops and then on a repetition of its first crop. The repeated crop retains the speaker's appearance while removing articulation. If the generator produces no speech, its empty transcript receives no evidence, does not vote, and cannot be selected. When USR\,2.0 replaces Auto-AVSR as the lip reader, only the hypotheses of $R$ change. The evidence terms are computed as above, and the audio-visual hypotheses for noisy input still come from Auto-AVSR, so the swap needs no retraining and no new setting.

Each term is normalized against its own baseline. Subtracting the same recognizer's log-likelihood of the same transcript under an uninformative input largely cancels the recognizer's language prior and the cost of the transcript's length. What remains is an estimate of the pointwise mutual information between the transcript and the input. Both terms are therefore log-likelihood ratios in the same unit, and we sum them without a modality weight: a modality that carries no word information gives a value near zero for every candidate and drops out. We sum over tokens rather than average, so that a transcript supported word by word outweighs one supported by a single confident word. Every window has the same length, which keeps this scale comparable across clips. The only global scale is the temperature $T$.

\Cref{tab:supp:evidence} reports both terms on the condition grid of \cref{sec:exp:adapt}. The audio evidence decreases as the damage increases. Its per-sample maximum over the candidates, which is the quantity quoted in \cref{sec:exp:adapt}, is high while speech remains audible, falls as the babble grows louder and covers more of the window, and vanishes when the whole clip is silenced. The visual evidence stays at a similar level in every condition, since the video is never corrupted.

\begin{table}[htbp]
  \centering
  \footnotesize
  \setlength{\tabcolsep}{10.0pt}
  \caption{\textbf{Evidence across the condition grid.} Mean audio evidence $e_A$ and visual evidence $e_R$ (in nats) of each condition of \cref{fig:adapt}. \emph{Sel.}: the selected output. \emph{Max}: the per-sample maximum over the candidates, which are the six configurations and the two text outputs $h_1$ and $\hat{y}_L$. 1w, 4w, clip: one word, four words or the whole window is affected.}
  \begin{tabular}{@{}lrrrrr@{}}
    \toprule
    & & \multicolumn{2}{c}{$e_A$} & \multicolumn{2}{c}{$e_R$} \\
    \cmidrule(lr){3-4} \cmidrule(l){5-6}
    Condition & $n$ & Sel. & Max & Sel. & Max \\
    \midrule
    Clean                    & 268 & 79.0 & 80.4 & 57.1 & 69.8 \\
    \midrule
    Babble $-5$\,dB, 1w      & 268 & 78.8 & 81.2 & 57.1 & 69.7 \\
    Babble $-5$\,dB, 4w      & 252 & 76.0 & 78.5 & 56.4 & 70.1 \\
    Babble $-5$\,dB, clip    & 268 & 55.2 & 58.7 & 53.6 & 69.6 \\
    \midrule
    Babble $-10$\,dB, 1w     & 268 & 78.1 & 81.1 & 57.1 & 69.7 \\
    Babble $-10$\,dB, 4w     & 252 & 70.1 & 73.5 & 54.5 & 70.1 \\
    Babble $-10$\,dB, clip   & 268 & 18.8 & 23.3 & 55.1 & 69.4 \\
    \midrule
    Babble $-20$\,dB, 1w     & 268 & 75.9 & 79.5 & 56.4 & 69.6 \\
    Babble $-20$\,dB, 4w     & 252 & 62.3 & 66.7 & 53.9 & 69.9 \\
    Babble $-20$\,dB, clip   & 268 & 0.2  & 3.3  & 55.8 & 69.3 \\
    \midrule
    Silence, 1w              & 268 & 74.8 & 77.0 & 57.0 & 69.6 \\
    Silence, 4w              & 252 & 60.9 & 64.1 & 55.6 & 69.8 \\
    Silence, clip            & 267 & 0.0  & 0.0  & 59.1 & 69.3 \\
    \bottomrule
  \end{tabular}
  \label{tab:supp:evidence}
\end{table}

\paragraph{Voters.} The language model reads the lip reader's full $N$-best list, so its transcript $\hat{y}_L$ already reflects information from $h_1$. Allowing both to vote would count that information twice. When $\hat{y}_L$ votes, we therefore retain the configuration that speaks $h_1$ over silence as a candidate but exclude it as a voter. The distance $d$ averages word error rate in both directions. Exact risk ties, which arise when configurations produce the same transcript, favor a configuration that keeps the audio and then one using more components. We did not tune this exclusion: it is a design rule that follows from the double-counting argument above. The only tuned parameter of the selection is the temperature, which we set to $T = 100$ on the validation grid.

\begin{table}[htbp]
  \centering
  \small
  \setlength{\tabcolsep}{4pt}
  \caption{\textbf{Configurations chosen by the selection} on the silenced cuts (pooled TalkVid and VoxMM) and on the noisy windows, in \% of samples. \emph{Muted}: the generator starts from silence instead of $\tilde{a}$.}
  \begin{tabular}{@{}lrrrrr@{}}
    \toprule
    & \multicolumn{3}{c}{Audio kept} & \multicolumn{2}{c}{Muted} \\
    \cmidrule(lr){2-4} \cmidrule(l){5-6}
    & $R{\rightarrow}L{\rightarrow}G$ & $R{\rightarrow}G$ & $G$ & $R{\rightarrow}L{\rightarrow}G$ & $R{\rightarrow}G$ \\
    \midrule
    Silenced (2,092) & 73.2 & 21.7 & 2.5 & 1.9 & 0.8 \\
    Noisy (1,365)    & 62.5 & 21.1 & 3.8 & 11.9 & 0.7 \\
    \bottomrule
  \end{tabular}
  \label{tab:supp:choices}
\end{table}

\paragraph{What the selection chooses.} On silenced cuts, the selection almost always keeps the audio, and for three quarters of the cuts it gives the generator the language model's transcript (\cref{tab:supp:choices}). On noisy windows it mutes the input more often, and the more so the stronger the interference: for 43\% of the windows under $-10$\,dB babble, 7\% under $-5$\,dB babble, and 4 to 5\% under sound effects or a competing speaker. On the condition grid, it keeps the audio for every clean clip but mutes 88\% of the clips fully covered by $-20$\,dB babble (\cref{fig:adapt}): of these 268 clips, it chooses $R \rightarrow L \rightarrow G$ over silence for 213 and $R \rightarrow G$ over silence for 23, and keeps the audio for the remaining 32. It never chooses $G$ over silence, the configuration with neither text nor audio, on the silenced cuts or noisy windows.

\begin{table}[htbp]
  \centering
  \footnotesize
  \setlength{\tabcolsep}{3.5pt}
  \caption{\textbf{Ablation of the selection rule.} \emph{Grid} and \emph{Noisy}: WER on the 3,419 grid samples of \cref{fig:adapt} and on the 1,365 noisy windows. \emph{Silenced}: exact span recovery on the 2,092 verified cuts. In parentheses: variant minus ours, bold where its 95\% interval excludes zero. The oracle picks the best configuration for each sample.}
  \begin{tabular}{@{}lr@{\,}lr@{\,}lr@{\,}l@{}}
    \toprule
    & \multicolumn{2}{c}{Grid $\downarrow$} & \multicolumn{2}{c}{Noisy $\downarrow$} & \multicolumn{2}{c}{Silenced $\uparrow$} \\
    \midrule
    Ours                         & 23.1 & & 27.2 & & 39.3 & \\
    \midrule
    All outputs vote             & 23.6 & ($+0.4$)          & 32.2 & ($\mathbf{+5.0}$) & 36.3 & ($\mathbf{-3.0}$) \\
    No evidence ($T \to \infty$) & 23.6 & ($\mathbf{+0.4}$) & 27.9 & ($\mathbf{+0.7}$) & 39.0 & ($-0.3$) \\
    Evidence only ($T \to 0$)    & 23.9 & ($\mathbf{+0.8}$) & 27.9 & ($\mathbf{+0.7}$) & 34.3 & ($\mathbf{-5.0}$) \\
    Without $e_A$                & 23.3 & ($+0.1$)          & 28.5 & ($\mathbf{+1.3}$) & 39.3 & ($+0.0$) \\
    Without $e_R$                & 23.5 & ($\mathbf{+0.3}$) & 27.4 & ($+0.3$)          & 38.9 & ($-0.4$) \\
    \midrule
    Oracle choice                & 20.0 & & 23.2 & & 44.3 & \\
    \bottomrule
  \end{tabular}
  \label{tab:supp:selection_ablation}
\end{table}

\paragraph{Ablation.} \Cref{tab:supp:selection_ablation} evaluates the selection rule on the condition grid of \cref{sec:exp:adapt}, the noisy windows, and the silenced cuts after removing one part at a time. Each part contributes significantly in at least one setting, and no ablated variant significantly improves on the full rule. The voter exclusion has the largest effect, on both noisy speech (CI $[3.7, 6.4]$ points) and silenced speech (CI $[1.2, 4.9]$). Selecting the highest-evidence output without a vote is worse in all three settings. Equal weighting ($T \to \infty$) costs less than a point on average, because the evidence changes the choice for only about one input in ten. On those inputs its effect is large (\cref{tab:supp:selection_diagnostics}b). Audio evidence matters on noisy speech and visual evidence on the grid, while on silenced speech removing either term has no significant effect. The oracle row shows the headroom that remains for a better rule.

\begin{table}[htbp]
  \centering
  \footnotesize
  \setlength{\tabcolsep}{2.5pt}
  \caption{\textbf{Selection diagnostics under Whisper scoring.} (a) Selection against the best fixed configuration chosen in hindsight for each condition. (b) Evidence weighting against equal weighting on the samples where their choices differ, with their WERs (E / eq.). $\Delta$: selection minus comparator. B/W: conditions significantly better or worse in (a), samples better or worse in (b).}
  \textbf{(a) Condition-wise hindsight rule}\\[2pt]
  \begin{tabular}{@{}lrrll@{}}
    \toprule
    Set & Sel. & Fixed & $\Delta$ [95\% CI] & B/W \\
    \midrule
    Noisy (1,365) & 27.2 & 28.7 & $-1.5$ $[-1.9,-1.0]$ & 4/0 \\
    Grid (3,419)  & 23.1 & 24.2 & $-1.0$ $[-1.3,-0.7]$ & 8/0 \\
    \bottomrule
  \end{tabular}\\[5pt]
  \textbf{(b) Disagreement subset}\\[2pt]
  \begin{tabular}{@{}lrlll@{}}
    \toprule
    Set & Changed & E / eq. & $\Delta$ [95\% CI] & B/W \\
    \midrule
    Noisy & 145 (10.6\%) & 40.6/47.3 & $-6.7$ $[-9.3,-3.9]$ & 80/29 \\
    Grid  & 304 (8.9\%)  & 35.0/39.8 & $-4.8$ $[-6.2,-3.5]$ & 155/77 \\
    \bottomrule
  \end{tabular}
  \label{tab:supp:selection_diagnostics}
\end{table}

\begin{table*}[t]
  \centering
  \footnotesize
  \setlength{\tabcolsep}{3.4pt}
  \caption{\textbf{Condition grid in full.} WER (\%) on the 3,419 samples of \cref{fig:adapt}. Babble or silence affects one word (1w), four words (4w) or the whole window (clip). \emph{Kept}: the generator starts from $\tilde{a}$. \emph{Muted}: a reference in which the damaged part of $\tilde{a}$ is replaced with silence. These rows are told where the damage is, unlike the selection, whose muted candidates start from a fully silent window. $R$ and $R \rightarrow L$ are scored on their text.}
  \begin{tabular}{@{}lrrrrrrrrrrrrrr@{}}
    \toprule
    & & \multicolumn{3}{c}{Babble $-5$\,dB} & \multicolumn{3}{c}{Babble $-10$\,dB} & \multicolumn{3}{c}{Babble $-20$\,dB} & \multicolumn{3}{c}{Silence} & \\
    \cmidrule(lr){3-5} \cmidrule(lr){6-8} \cmidrule(lr){9-11} \cmidrule(lr){12-14}
    Method & Clean & 1w & 4w & clip & 1w & 4w & clip & 1w & 4w & clip & 1w & 4w & clip & Avg \\
    \midrule
    Samples                                  & 268 & 268 & 252 & 268 & 268 & 252 & 268 & 268 & 252 & 268 & 268 & 252 & 267 & 3,419 \\
    \midrule
    Selection (ours)                         & 10.1 & 10.4 & 13.1 & 27.1 & 11.2 & 18.1 & 43.3 & 12.3 & 22.2 & 49.8 & 12.2 & 21.3 & 48.9 & \textbf{23.1} \\
    \midrule
    $R \rightarrow L \rightarrow G$, kept    & 10.3 & 11.4 & 14.6 & 28.4 & 12.2 & 20.3 & 47.5 & 13.7 & 27.3 & 75.4 & 13.2 & 22.4 & 50.0 & 26.8 \\
    $R \rightarrow L \rightarrow G$, muted   & 50.0 & 13.2 & 22.4 & 50.0 & 13.2 & 22.4 & 50.0 & 13.2 & 22.4 & 50.0 & 13.2 & 22.4 & 50.0 & 30.3 \\
    $R \rightarrow G$, kept                  & 10.2 & 11.6 & 14.9 & 29.0 & 12.9 & 19.0 & 47.9 & 14.1 & 25.5 & 71.3 & 13.6 & 22.9 & 49.3 & 26.4 \\
    $R \rightarrow G$, muted                 & 49.5 & 13.6 & 22.9 & 49.5 & 13.6 & 22.9 & 49.5 & 13.6 & 22.9 & 49.5 & 13.6 & 22.9 & 49.3 & 30.4 \\
    $G$ without text, kept                   & 10.2 & 12.7 & 16.2 & 42.0 & 14.4 & 26.1 & 81.4 & 17.1 & 39.0 & 114.9 & 15.9 & 38.6 & 100.1 & 40.8 \\
    $G$ without text, muted                  & 100.1 & 15.9 & 38.6 & 100.1 & 15.9 & 38.6 & 100.1 & 15.9 & 38.6 & 100.1 & 15.9 & 38.6 & 100.1 & 55.6 \\
    \midrule
    $R$ (text)                               & 48.8 & 48.8 & 49.9 & 48.8 & 48.8 & 49.9 & 48.8 & 48.8 & 49.9 & 48.8 & 48.8 & 49.9 & 48.8 & 49.1 \\
    $R \rightarrow L$ (text)                 & 16.9 & 18.4 & 24.6 & 38.7 & 20.5 & 32.6 & 46.4 & 22.8 & 38.9 & 47.7 & 17.5 & 30.5 & 47.9 & 31.0 \\
    \bottomrule
  \end{tabular}
  \label{tab:supp:grid}
\end{table*}

\paragraph{Condition grid in full.} \Cref{tab:supp:grid} gives the WER behind \cref{fig:adapt} for the selection, every fixed configuration and the two text outputs. The averages quoted in \cref{sec:exp:adapt} are its last column, with the input audio kept for the fixed configurations. No fixed configuration is best in every condition. Those that keep the audio lead while speech remains audible, but they fail under whole-clip $-20$\,dB babble, where the same configurations over silence are far better. The selection adapts to both cases: it keeps the audio while speech is audible and switches to the configurations over silence when the babble covers the whole clip. Under whole-clip $-20$\,dB babble it stays within about a point of the muted configurations and of the lip reader's own text, without being told the noise level. The muted rows of \cref{tab:supp:grid} are reference rows that are given the damaged region and replace only that region with silence, which is why they repeat the silence columns. The selection does not receive the damaged region. Its muted candidates discard the whole input track, and it chooses between them and the candidates that keep the audio using the evidence alone. Even so, it matches or improves on the muted rows in every condition except whole-clip $-20$\,dB babble, where it is within half a point.

\paragraph{Against condition-wise fixed choices.} The fixed configurations in \cref{tab:supp:grid} use one choice for every input. For a stronger retrospective comparator, we choose the lowest-WER configuration separately for each condition using that condition's test outcomes. It therefore knows the condition and the results, and its muted configurations also know the damaged region, but it still uses one configuration for every sample in that condition. The per-sample oracle of \cref{tab:supp:selection_ablation} is a different comparator. Under Whisper scoring, selection outperforms this condition-wise rule on both the noisy set and the condition grid (\cref{tab:supp:selection_diagnostics}a). Five configurations lead across the grid's 13 conditions, including $G$ alone on clean input and muted configurations under $-20$\,dB babble.

\paragraph{When evidence changes the choice.} Evidence changes the selected configuration for only a small fraction of inputs. On those disagreement subsets, its choice has substantially lower WER than equal-weight voting (\cref{tab:supp:selection_diagnostics}b). This conditional analysis explains why evidence can matter for individual inputs while the aggregate difference in \cref{tab:supp:selection_ablation} is modest.

\section{Additional results}
\label{sec:supp:results}

We first examine paired comparisons with the baselines, then test whether the noisy-speech results hold across interference sources and test sets. The remaining analyses probe the generator, the scoring protocol, examples of the recovered speech, gaps that are not digital silence, and what the language model takes from the lip reader.

\subsection{Comparison with the baselines}
\label{sec:supp:significance}

The main paper reports method scores. Here we give the paired differences and 95\% intervals behind those comparisons (\cref{sec:supp:scoring}). We begin with the pooled silenced and noisy evaluations, then examine the hard set on the inputs each baseline can process.

\begin{table}[htbp]
  \centering
  \footnotesize
  \setlength{\tabcolsep}{2pt}
  \caption{\textbf{Comparison with the baselines}, pooled over TalkVid and VoxMM. $\Delta$: ours minus the baseline, in points, with its 95\% interval, bold where the interval excludes zero. Positive $\Delta$ favors our method in (a), negative $\Delta$ in (b). \emph{Tied}: noise conditions in which the difference is not significant. $^\dagger$On the windows the method covers.}
  \label{tab:supp:ci}
  \textbf{(a)} Silenced speech, verified cuts\\[3pt]
  \begin{tabular*}{\columnwidth}{@{}l@{\extracolsep{\fill}}r@{\extracolsep{0pt}\hspace{4pt}}l@{\extracolsep{\fill}}r@{\extracolsep{0pt}\hspace{4pt}}l@{}}
    \toprule
    & \multicolumn{2}{c}{Exact recovery} & \multicolumn{2}{c}{Word accuracy} \\
    \cmidrule{2-3} \cmidrule{4-5}
    Method & \% & $\Delta$ {\scriptsize[95\% CI]} & \% & $\Delta$ {\scriptsize[95\% CI]} \\
    \midrule
    Ours (Auto-AVSR as $R$)          & 39.3 &                                         & 65.7 & \\
    \midrule
    Auto-AVSR~\cite{ma2023auto}      & 30.7 & $\mathbf{+8.6}$ {\scriptsize$[6.4, 11.0]$}  & 59.7 & $\mathbf{+6.0}$ {\scriptsize$[4.7, 7.2]$} \\
    USR\,2.0~\cite{haliassos2026pay} & 35.8 & $+3.5$ {\scriptsize$[-0.6, 8.0]$}           & 64.9 & $+0.8$ {\scriptsize$[-2.3, 4.0]$} \\
    Llama-AVSR~\cite{10889251}       & 30.4 & $\mathbf{+8.9}$ {\scriptsize$[3.5, 14.6]$}  & 58.5 & $\mathbf{+7.1}$ {\scriptsize$[1.8, 11.9]$} \\
    LipVoicer~\cite{yemini2024lipvoicer} & 28.3 & $\mathbf{+11.0}$ {\scriptsize$[7.0, 15.9]$} & 58.6 & $\mathbf{+7.0}$ {\scriptsize$[3.8, 10.1]$} \\
    VoiceCraft~\cite{peng-etal-2024-voicecraft} ($L$) & 32.3 & $\mathbf{+7.0}$ {\scriptsize$[5.5, 8.7]$} & 59.8 & $\mathbf{+5.8}$ {\scriptsize$[4.8, 6.9]$} \\
    F5-TTS~\cite{chen-etal-2025-f5} ($L$) & 35.8 & $\mathbf{+3.5}$ {\scriptsize$[2.4, 4.7]$} & 62.2 & $\mathbf{+3.5}$ {\scriptsize$[2.2, 4.7]$} \\
    AlignDiT~\cite{Choi_2025} ($L$)  & 35.7 & $\mathbf{+3.6}$ {\scriptsize$[2.3, 4.9]$}   & 62.8 & $\mathbf{+2.9}$ {\scriptsize$[1.7, 4.0]$} \\
    \bottomrule
  \end{tabular*}

  \vspace{8pt}
  \textbf{(b)} Noisy speech, average over the five conditions\\[3pt]
  \begin{tabular*}{\columnwidth}{@{}l@{\extracolsep{\fill}}r@{\extracolsep{0pt}\hspace{4pt}}l@{\extracolsep{\fill}}l@{}}
    \toprule
    Method & WER & $\Delta$ {\scriptsize[95\% CI]} & Tied \\
    \midrule
    Ours (Auto-AVSR as $R$)    & 27.2 &                                            & \\
    \midrule
    Noisy input                & 69.0 & $\mathbf{-41.8}$ {\scriptsize$[-46.1, -38.2]$} & -- \\
    Auto-AVSR (V)              & 49.4 & $\mathbf{-22.2}$ {\scriptsize$[-26.4, -17.6]$} & -- \\
    Auto-AVSR (AV)             & 70.6 & $\mathbf{-43.4}$ {\scriptsize$[-46.1, -40.6]$} & -- \\
    USR\,2.0 (V)               & 47.4 & $\mathbf{-20.2}$ {\scriptsize$[-25.2, -15.5]$} & babble $-10$ \\
    USR\,2.0 (AV)              & 52.0 & $\mathbf{-24.8}$ {\scriptsize$[-28.2, -21.7]$} & babble $-5$ \\
    Llama-AVSR (V)             & 58.8 & $\mathbf{-31.6}$ {\scriptsize$[-39.4, -23.7]$} & -- \\
    Llama-AVSR (AV)            & 47.9 & $\mathbf{-20.6}$ {\scriptsize$[-23.9, -17.4]$} & -- \\
    DeepFilterNet3             & 72.2 & $\mathbf{-44.9}$ {\scriptsize$[-49.8, -41.1]$} & -- \\
    Sidon                      & 84.4 & $\mathbf{-57.1}$ {\scriptsize$[-61.6, -53.4]$} & -- \\
    SAM Audio, text prompt     & 67.1 & $\mathbf{-39.9}$ {\scriptsize$[-44.7, -35.7]$} & -- \\
    SAM Audio, visual prompt   & 75.4 & $\mathbf{-48.2}$ {\scriptsize$[-52.6, -43.7]$} & -- \\
    VisualVoice$^\dagger$      & 40.1 & $\mathbf{-12.7}$ {\scriptsize$[-16.0, -9.8]$}  & other talker \\
    LipVoicer                  & 52.9 & $\mathbf{-25.7}$ {\scriptsize$[-30.2, -20.9]$} & -- \\
    AlignDiT ($R$)$^\dagger$   & 50.5 & $\mathbf{-23.4}$ {\scriptsize$[-27.7, -18.9]$} & -- \\
    \bottomrule
  \end{tabular*}
\end{table}

\paragraph{Silenced speech.} With Auto-AVSR as the lip reader, our method significantly exceeds every baseline except USR\,2.0, with which it is statistically tied on the pooled cuts (\cref{tab:supp:ci}a). The comparison varies by test set: our lead over USR\,2.0 is 6.9 points on TalkVid (CI $[0.5, 13.2]$) but 1.2 on VoxMM (CI $[-4.6, 7.0]$). On VoxMM, the difference from Llama-AVSR also includes zero ($+6.2$, CI $[-0.4, 11.9]$).

USR\,2.0 is a stronger lip reader than Auto-AVSR. Substituting it for Auto-AVSR raises our exact recovery from 39.3\% to 43.8\% ($+4.5$, CI $[1.1, 7.7]$), 8.0 points (CI $[5.3, 11.0]$) above USR\,2.0 alone.

\paragraph{Noisy speech.} Averaged over the five conditions, our method has significantly lower WER than every baseline (\cref{tab:supp:ci}b). No baseline is significantly better on any pooled condition, although three are tied with it on one condition each.

Which baseline comes closest to our method depends on the type of interference. VisualVoice matches our competing-speaker result on the windows it covers ($+1.6$, CI $[-0.8, 4.0]$) and is better on VoxMM alone ($+3.9$, CI $[0.8, 6.9]$). USR\,2.0 using only the lips ties with us under $-10$\,dB babble ($-3.5$, CI $[-7.5, 0.6]$), where audio is least informative. USR\,2.0 using lips and audio ties under $-5$\,dB babble ($-0.6$, CI $[-4.5, 2.6]$), where much speech remains audible.

Replacing our lip reader with USR\,2.0 reduces average WER from 27.2\% to 25.5\% ($-1.7$, CI $[-3.1, -0.2]$). The babble comparison is repeated with other recordings in \cref{sec:supp:babble}.

We also run SAM Audio on the pooled TalkVid and VoxMM noisy samples, using ``speech'' as the text prompt or the masked target-speaker video as the visual prompt. Both variants have substantially higher average WER than our method with either output transcriber (\cref{tab:supp:ci}b and \cref{tab:supp:parakeet_noisy}). SAM Audio is also evaluated on the hard set in \cref{tab:supp:hard}.

\begin{table}[htbp]
  \centering
  \footnotesize
  \setlength{\tabcolsep}{0.6pt}
  \caption{\textbf{Hard set: coverage and differences.} $n$: cuts or windows a method covers. $\Delta$: ours minus the baseline on those, with its 95\% interval over scenes, bold where it excludes zero. Positive $\Delta$ favors our method for silenced speech, negative $\Delta$ for noisy speech.}
  \begin{tabular}{@{}lrcrc@{}}
    \toprule
    & \multicolumn{2}{c}{Silenced (exact)} & \multicolumn{2}{c}{Noisy (WER)} \\
    \cmidrule(lr){2-3} \cmidrule(l){4-5}
    Method & $n$ & $\Delta$ & $n$ & $\Delta$ \\
    \midrule
    Damaged input            & --  & --                                       & 220 & $\mathbf{-24.6}$ {\scriptsize$[-32.6, -17.6]$} \\
    Auto-AVSR (V / AV)       & 116 & $\mathbf{+17.2}$ {\scriptsize$[11.1, 22.9]$} & 210 & $\mathbf{-42.5}$ {\scriptsize$[-49.2, -34.7]$} \\
    USR\,2.0                 & 83  & $\mathbf{+10.8}$ {\scriptsize$[2.8, 17.6]$}  & 160 & $\mathbf{-27.4}$ {\scriptsize$[-34.1, -20.5]$} \\
    Llama-AVSR               & 83  & $+9.6$ {\scriptsize$[-1.8, 18.1]$}           & 160 & $\mathbf{-31.4}$ {\scriptsize$[-41.5, -22.6]$} \\
    SAM Audio, text          & --  & --                                       & 220 & $\mathbf{-24.1}$ {\scriptsize$[-29.2, -18.3]$} \\
    SAM Audio, video         & --  & --                                       & 220 & $\mathbf{-34.3}$ {\scriptsize$[-41.9, -26.4]$} \\
    VisualVoice              & --  & --                                       & 90  & $-8.8$ {\scriptsize$[-22.9, 2.0]$} \\
    LipVoicer                & 111 & $\mathbf{+16.2}$ {\scriptsize$[9.5, 22.5]$}  & 210 & $\mathbf{-64.3}$ {\scriptsize$[-80.8, -46.0]$} \\
    VoiceCraft ($L$)         & 116 & $\mathbf{+6.0}$ {\scriptsize$[0.9, 11.8]$}   & --  & -- \\
    F5-TTS ($L$)             & 116 & $+2.6$ {\scriptsize$[-1.1, 7.3]$}            & --  & -- \\
    AlignDiT ($L$ / $R$)     & 110 & $+3.6$ {\scriptsize$[-0.8, 8.8]$}            & 165 & $\mathbf{-68.8}$ {\scriptsize$[-86.6, -47.7]$} \\
    \bottomrule
  \end{tabular}
  \label{tab:supp:hard}
\end{table}

\paragraph{Hard set.} \Cref{tab:supp:hard} gives the differences behind \cref{tab:hard}, comparing our method with each baseline on the cuts or windows that baseline covers. For silenced speech, our advantage is significant over Auto-AVSR, USR\,2.0 on the cuts it covers, LipVoicer, and VoiceCraft. Llama-AVSR trails on the same cuts as USR\,2.0, but its interval includes zero.

F5-TTS and AlignDiT, both given our language model's transcript and the span location, are statistically tied with us. Much of the gain on this set comes from the language model (\cref{tab:ablations}a), whose transcript these editors also use. For noisy speech, our method significantly improves on every baseline except VisualVoice. Its face detector succeeds on fewer than half of the windows, and the two methods tie on that subset.

\subsection{Babble from other recordings}
\label{sec:supp:babble}

The main evaluation uses DEMAND cafeteria babble on microphone channels held out from training (\cref{sec:supp:corruptions}). To examine other babble, we test the same speech windows with two further sources. One is DEMAND's restaurant scene, recorded in a different room and session. The other sums speech from six talkers in six held-out test videos, excluding the source video of each window. Neither source shares a recording, scene, or speaker with training. We replace the babble in all 546 samples at both levels from \cref{tab:noisy}, keeping the video, clean speech, generator text, sampling seed, and mixing code fixed. Only the interference and its mixing level change.

\paragraph{Matched input WER.} At a fixed signal-to-noise ratio, different babble recordings need not mask speech equally. The cafeteria recording has 39\% of its energy between 300 and 3,400\,Hz, compared with 69\% for the restaurant recording. At the nominal $-5$\,dB level, restaurant babble and the six-talker mixture raise noisy-input WER from 52.9\% to above 90\%. A comparison at the same ratio would therefore confound babble source with input difficulty. Instead, we choose ratios that match Whisper's noisy-input WER under cafeteria babble: $-5.4$ and $-1.4$\,dB for the restaurant, and $-4.6$ and $-0.4$\,dB for the six talkers. We fix these ratios using only the noisy inputs, before running any restoration method. At each level, input WER then differs from the cafeteria condition by at most 1.3 points (\cref{tab:supp:babble}, first row).

\paragraph{Methods.} For each new mixture, we regenerate the three configurations that use the input audio and rerun selection with the same rule and temperature, recomputing only its input-dependent evidence. We also rerun DeepFilterNet3, Sidon, and VisualVoice with their original code. LipVoicer and AlignDiT do not use the input audio (\cref{tab:supp:inputs}), so their outputs are identical across babble sources. As a reproducibility check, rerunning our cafeteria-babble outputs on a second cluster changes WER by no more than 0.1 points.

\begin{table*}[t]
  \centering
  \footnotesize
  \setlength{\tabcolsep}{4pt}
  \caption{\textbf{Babble from other recordings.} WER (\%) on the 546 babble samples of \cref{tab:noisy} under the cafeteria babble of the main experiments and under two babble sources that share no recording and no speaker with training, mixed to the same input WER. $\Delta$: the selection minus the method on the same windows, with its 95\% interval for the new sources. Every interval excludes zero. $^\ast$Does not read the input audio. $^\dagger$On the windows the method covers (VisualVoice 498, AlignDiT 526).}
  \begin{tabular}{@{}lrrr@{\hspace{16pt}}rll@{}}
    \toprule
    & \multicolumn{3}{c}{WER} & \multicolumn{3}{c}{$\Delta$ (ours $-$ method)} \\
    \cmidrule(r{16pt}){2-4} \cmidrule{5-7}
    Method & Cafeteria & Restaurant & Six talkers & Cafeteria & Restaurant & Six talkers \\
    \midrule
    Noisy input                    & 72.3 & 72.6 & 72.9 & $-36.3$ & $-36.6$ {\scriptsize$[-43.2, -31.0]$} & $-35.7$ {\scriptsize$[-42.4, -29.8]$} \\
    DeepFilterNet3                 & 81.6 & 81.1 & 89.4 & $-45.6$ & $-45.1$ {\scriptsize$[-51.0, -39.9]$} & $-52.2$ {\scriptsize$[-58.9, -46.2]$} \\
    Sidon                          & 96.2 & 93.9 & 99.7 & $-60.2$ & $-57.9$ {\scriptsize$[-63.9, -52.7]$} & $-62.5$ {\scriptsize$[-70.3, -55.6]$} \\
    VisualVoice$^\dagger$          & 61.1 & 58.1 & 60.3 & $-24.7$ & $-21.4$ {\scriptsize$[-27.4, -16.1]$} & $-22.5$ {\scriptsize$[-28.6, -17.2]$} \\
    LipVoicer$^\ast$               & 52.9 & 52.9 & 52.9 & $-16.9$ & $-16.9$ {\scriptsize$[-20.4, -12.9]$} & $-15.7$ {\scriptsize$[-19.7, -11.3]$} \\
    AlignDiT ($R$)$^{\ast\dagger}$ & 50.5 & 50.5 & 50.5 & $-14.7$ & $-14.6$ {\scriptsize$[-18.2, -11.4]$} & $-13.4$ {\scriptsize$[-16.8, -10.3]$} \\
    \midrule
    Ours, $G$ alone                       & 64.4 & 62.3 & 75.3 & & & \\
    Ours, $R \rightarrow G$               & 38.3 & 39.4 & 42.4 & & & \\
    Ours, $R \rightarrow L \rightarrow G$ & 39.1 & 38.1 & 41.0 & & & \\
    Ours, selection                       & \textbf{36.0} & \textbf{36.0} & \textbf{37.2} & & & \\
    \bottomrule
  \end{tabular}
  \label{tab:supp:babble}
\end{table*}

\paragraph{Results.} Our method's WER remains comparable across the cafeteria and the two independent babble sources (\cref{tab:supp:babble}). Relative to cafeteria babble, the restaurant difference is $+0.0$ (CI $[-1.7, 1.7]$) and the six-talker difference is $+1.2$ (CI $[-0.1, 2.5]$). Neither is significant, also when the two levels are considered separately. Our method remains significantly ahead of every baseline evaluated here for both new sources, with similar margins and the same baseline ordering. Under restaurant babble, none of its fixed configurations changes significantly either.

The six-talker mixture is harder for several audio-processing methods despite the matched input WER. DeepFilterNet3 and Sidon both lose significantly. Among our fixed configurations, $G$ alone is affected most, whereas the configurations given a transcript rise far less. The selection limits the effect by muting the input on 30\% of these windows, compared with 25\% under cafeteria babble, and hence relying more on the lips when the audio is harder to restore. Its margin over DeepFilterNet3 grows as a result.

\paragraph{Same nominal ratio.} We also test restaurant babble at the main experiment's nominal $-10$ and $-5$\,dB. This is a harder input condition: noisy-input WER increases from 72.3\% to 95.0\%. Output WER rises for every method, including $R \rightarrow L \rightarrow G$ at 51.0\% and DeepFilterNet3 at 95.0\%. Their relative difference remains comparable to that under cafeteria babble ($-44.0$, CI $[-52.4, -36.7]$, versus $-42.4$).

\subsection{Noisy speech on each test set}
\label{sec:supp:noisy_sets}

\begin{table*}[t]
  \centering
  \small
  \setlength{\tabcolsep}{4.5pt}
  \caption{\textbf{Noisy speech on each test set.} WER (\%, lower is better) of whole windows on TalkVid (570 windows) and VoxMM (795 windows). \Cref{tab:noisy} pools the two sets. $^\dagger$Only windows on which the method's face preprocessing succeeds (VisualVoice 490 and 755, AlignDiT 555 and 760).}
  \begin{tabular}{@{}lrrrrrr@{\hspace{14pt}}rrrrrr@{}}
    \toprule
    & \multicolumn{6}{c}{TalkVid} & \multicolumn{6}{c}{VoxMM} \\
    \cmidrule(r{14pt}){2-7} \cmidrule{8-13}
    Method & B$-10$ & B$-5$ & seen & unseen & talker & Avg & B$-10$ & B$-5$ & seen & unseen & talker & Avg \\
    \midrule
    Noisy input                    & 87.2 & 49.5 & 42.1 & 46.0 & 102.9 & 65.5 & 95.0 & 55.3 & 51.7 & 49.4 & 106.2 & 71.5 \\
    \midrule
    Auto-AVSR (V)                  & 29.3 & 29.3 & 29.3 & 29.3 & 29.3  & 29.3 & 63.8 & 63.8 & 63.8 & 63.8 & 63.8  & 63.8 \\
    Auto-AVSR (AV)                 & 66.0 & 32.6 & 55.8 & 47.0 & 84.0  & 57.1 & 93.6 & 58.6 & 78.0 & 71.0 & 100.4 & 80.3 \\
    USR\,2.0 (V)                   & 30.4 & 30.4 & 30.4 & 30.4 & 30.4  & 30.4 & 59.7 & 59.7 & 59.7 & 59.7 & 59.7  & 59.7 \\
    USR\,2.0 (AV)                  & 46.6 & 22.9 & 28.2 & 21.6 & 104.7 & 44.8 & 73.8 & 33.0 & 40.6 & 34.3 & 104.0 & 57.1 \\
    Llama-AVSR (V)                 & 35.2 & 35.2 & 35.2 & 35.2 & 35.2  & 35.2 & 75.8 & 75.8 & 75.8 & 75.8 & 75.8  & 75.8 \\
    Llama-AVSR (AV)                & 37.2 & 21.2 & 25.2 & 19.1 & 85.0  & 37.5 & 72.6 & 43.7 & 34.6 & 31.0 & 94.3  & 55.3 \\
    \midrule
    DeepFilterNet3                 & 87.8 & 64.0 & 48.7 & 41.0 & 101.4 & 68.6 & 95.9 & 75.4 & 53.6 & 45.0 & 103.8 & 74.7 \\
    Sidon                          & 98.9 & 88.6 & 57.2 & 61.6 & 102.0 & 81.7 & 99.8 & 96.1 & 64.6 & 67.0 & 103.9 & 86.3 \\
    VisualVoice$^\dagger$          & 69.8 & 31.3 & 24.9 & 24.8 & 15.7  & 33.3 & 87.4 & 48.6 & 31.1 & 32.0 & 24.0  & 44.6 \\
    LipVoicer                      & 33.2 & 33.2 & 33.2 & 33.2 & 33.2  & 33.2 & 67.0 & 67.0 & 67.0 & 67.0 & 67.0  & 67.0 \\
    AlignDiT ($R$)$^\dagger$       & 30.5 & 30.5 & 30.5 & 30.5 & 30.5  & 30.5 & 65.0 & 65.0 & 65.0 & 65.0 & 65.0  & 65.0 \\
    Ours (Auto-AVSR as $R$)        & 24.1 & 17.8 & 14.8 & 13.9 & 14.1  & 16.9 & 58.2 & 35.5 & 27.2 & 24.2 & 27.8  & 34.6 \\
    Ours (USR\,2.0 as $R$)         & 23.4 & 16.3 & 14.6 & 13.5 & 13.7  & 16.3 & 52.0 & 34.0 & 24.8 & 23.0 & 26.5  & 32.1 \\
    \midrule
    Clean recording                & 8.5  & 8.5  & 8.5  & 8.5  & 8.5   & 8.5  & 13.0 & 13.0 & 13.0 & 13.0 & 13.0  & 13.0 \\
    \bottomrule
  \end{tabular}
  \label{tab:supp:noisy_sets}
\end{table*}

\Cref{tab:supp:noisy_sets} disaggregates \cref{tab:noisy} by test set. WER is higher on VoxMM even for the clean recordings, and the lip readers alone are about twice as wrong there as on TalkVid. These differences help explain why every method performs worse on VoxMM.

Our method has the lowest average WER on both sets. The two sets differ most under $-10$\,dB babble, the condition in which performance relies most on the lip reader. There, replacing Auto-AVSR with USR\,2.0 in our pipeline hardly changes the result on TalkVid but lowers the WER by six points on VoxMM.

\subsection{Silenced speech by span length}
\label{sec:supp:span_length}

\begin{figure*}[t]
  \centering
  \includegraphics[width=\linewidth]{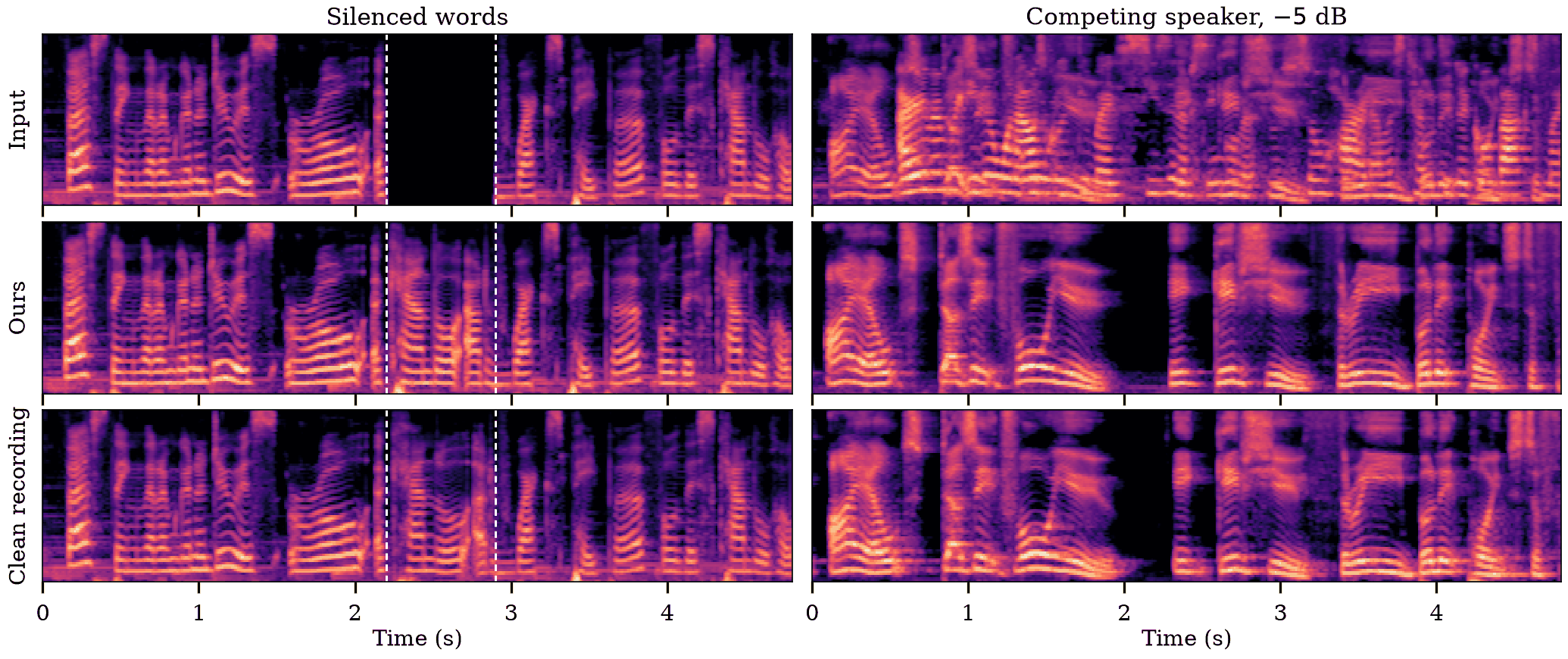}
  \caption{\textbf{Two test examples} as mel spectrograms: the input, the output of $R \rightarrow L \rightarrow G$, and the clean recording. \emph{Left:} the cut of \cref{fig:supp:prompt}, with dashed lines around the silenced words ``candle out''. \emph{Right:} a competing speaker at $-5$\,dB. Whisper transcribes the input as the other speaker's sentence, and the output as the reference, ``IELTS does it for you. It gives you three bullet points to''.}
  \label{fig:supp:examples}
\end{figure*}

\begin{figure}[htbp]
  \centering
  \includegraphics[width=\linewidth]{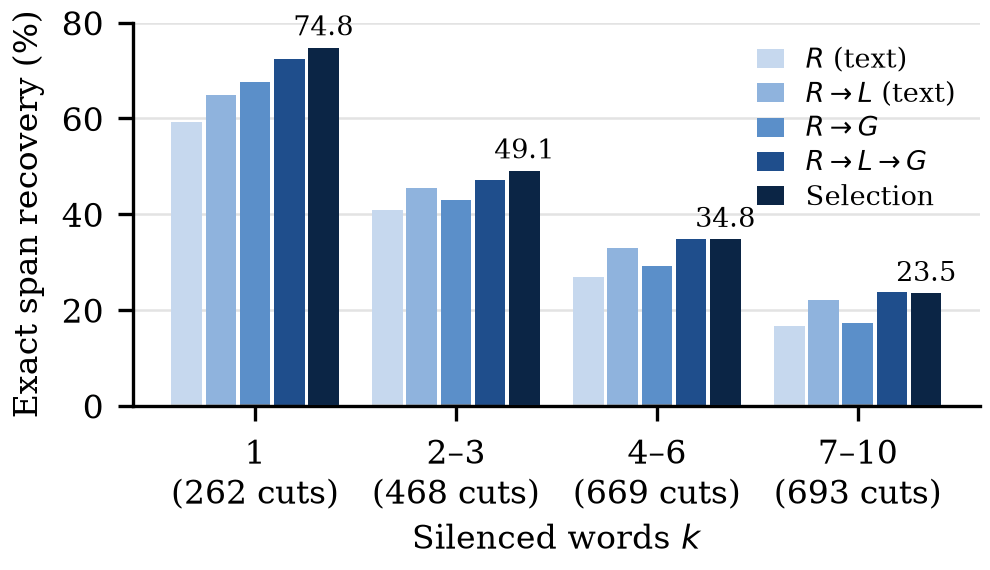}
  \caption{\textbf{Exact span recovery by number of silenced words}, on the verified TalkVid and VoxMM cuts. $R$ and $R \rightarrow L$ are scored on their text, the other three on the generated speech. Auto-AVSR is the lip reader.}
  \label{fig:supp:span_length}
\end{figure}

\Cref{fig:supp:span_length} breaks the silenced results down by the number of silenced words. Every configuration recovers fewer spans as the span grows. The language model improves on the lip reader, and the generator speaking its transcript improves further at every length, from 59.2\% for the lip reader to 72.5\% for single words and from 16.7\% to 23.7\% for seven to ten words. The generator contributes most for single words: there, speaking the lip reader's best hypothesis ($R \rightarrow G$) already recovers clearly more spans than that hypothesis itself, whereas for seven to ten words the two are within a point. The selection adds about two points to $R \rightarrow L \rightarrow G$ for spans of up to three words and matches it for longer ones.

\subsection{Generator controls}
\label{sec:supp:checks}

The following controls examine how the generator responds to the supplied text and video.

\paragraph{The generator says what it is given.} To isolate the generator's contribution, we give $G$ the window's reference transcript in place of $L$'s transcript. Whisper then recovers the silenced span exactly in more than 95\% of the verified cuts of every set, including the hard set (\cref{tab:supp:truetext}a), and this holds for spans of five to ten words (95.8\% on TalkVid and VoxMM). Given the same reference transcript, every speech editor recovers fewer spans on every set, although the editors are also given the span's location. $G$ also remains the highest on the 108 hard-set cuts that every method covers. These oracle-text results show that the generator usually places and speaks the supplied words intelligibly, even when the mouth is difficult to read, so most remaining errors of the full pipeline arise before generation. They also help explain why a stronger lip reader improves results without retraining (\cref{sec:supp:significance}).

The generator also follows the supplied transcript under noise (\cref{tab:supp:truetext}b). Given the noisy input and the reference transcript, $G$ lowers the WER of our selection on every set, by about half when pooled, and on TalkVid and VoxMM it comes within about three points of the clean recordings (\cref{tab:supp:noisy_sets}). The hard set shows the same pattern, with the WER halving.

\begin{table}[htbp]
  \centering
  \footnotesize
  \setlength{\tabcolsep}{4pt}
  \caption{\textbf{Given the true transcript.} (a) Exact span recovery (\%) on the verified silenced cuts when each method receives the reference transcript. (b) Whole-window WER (\%) on the noisy samples: the input, our selection, and $G$ given the reference transcript. $^\dagger$On the 110 cuts the method covers.}
  \textbf{(a) Silenced speech}\\[2pt]
  \begin{tabular}{@{}lrrrr@{}}
    \toprule
    Method & TalkVid & VoxMM & Pooled & Hard \\
    \midrule
    Ours, $G$                                   & \textbf{98.1} & \textbf{95.5} & \textbf{96.6} & \textbf{95.7} \\
    F5-TTS~\cite{chen-etal-2025-f5}             & 96.2 & 92.8 & 94.2 & 90.5 \\
    AlignDiT~\cite{Choi_2025}                   & 94.3 & 88.4 & 90.9 & 92.7$^\dagger$ \\
    VoiceCraft~\cite{peng-etal-2024-voicecraft} & 86.7 & 82.5 & 84.3 & 87.9 \\
    \bottomrule
  \end{tabular}\\[5pt]
  \textbf{(b) Noisy speech}\\[2pt]
  \begin{tabular}{@{}lrrr@{}}
    \toprule
    Set & Input & Selection & True text \\
    \midrule
    TalkVid (570)  & 65.5 & 16.9 & 11.3 \\
    VoxMM (795)    & 71.5 & 34.6 & 16.5 \\
    Pooled (1,365) & 69.0 & 27.2 & 14.3 \\
    Hard (220)     & 54.1 & 29.6 & 15.2 \\
    \bottomrule
  \end{tabular}
  \label{tab:supp:truetext}
\end{table}

\begin{table*}[t]
  \centering
  \footnotesize
  \setlength{\tabcolsep}{5pt}
  \caption{\textbf{What the generator takes from the lips.} Only the video given to $G$ changes: the correct video (\emph{Corr.}), a frozen mouth, or mismatched lips. \emph{Same transcript}: $G$ is given the transcript obtained from the correct lips ($R \rightarrow L \rightarrow G$ in (a), $R \rightarrow G$ in (b)). The frozen and mismatched columns give the difference to the correct video, in points, with its paired 95\% interval.}
  \label{tab:supp:visual_intervention}
  \begin{tabular}{@{}l r @{\hspace{18pt}} r l l @{\hspace{18pt}} r l l@{}}
    \toprule
    & & \multicolumn{3}{c}{$G$ alone} & \multicolumn{3}{c}{Same transcript} \\
    \cmidrule(r{14pt}){3-5} \cmidrule{6-8}
    & $n$ & Corr. & Frozen mouth & Mismatched lips & Corr. & Frozen mouth & Mismatched lips \\
    \midrule
    \multicolumn{8}{@{}l}{\textbf{(a)} Silenced speech: exact span recovery (\%, higher is better) by number of silenced words} \\[2pt]
    All cuts     & 1,877 & 7.1  & $-3.6$ {\scriptsize$[-4.9, -2.5]$}  & $-4.6$ {\scriptsize$[-6.1, -3.4]$}   & 38.8 & $-2.2$ {\scriptsize$[-3.2, -1.4]$} & $-6.3$ {\scriptsize$[-9.1, -4.3]$} \\
    1--3 words   & 657   & 19.8 & $-9.9$ {\scriptsize$[-13.4, -6.8]$} & $-12.8$ {\scriptsize$[-16.6, -9.3]$} & 56.6 & $-3.2$ {\scriptsize$[-4.5, -1.8]$} & $-4.7$ {\scriptsize$[-6.3, -3.2]$} \\
    4--6 words   & 598   & 0.5  & $-0.5$ {\scriptsize$[-1.2, 0.0]$}   & $-0.5$ {\scriptsize$[-1.2, 0.0]$}    & 34.4 & $-1.8$ {\scriptsize$[-3.6, -0.4]$} & $-7.7$ {\scriptsize$[-11.7, -4.8]$} \\
    7--10 words  & 622   & 0.0  & $0.0$                               & $0.0$                                & 24.1 & $-1.6$ {\scriptsize$[-2.9, -0.5]$} & $-6.6$ {\scriptsize$[-11.5, -3.1]$} \\
    \midrule
    \multicolumn{8}{@{}l}{\textbf{(b)} Severe noise: whole-window WER (\%, lower is better) by affected part and babble level} \\[2pt]
    All samples        & 914 & 66.5  & $+10.4$ {\scriptsize$[6.2, 15.0]$}  & $+16.9$ {\scriptsize$[13.6, 21.1]$} & 40.9 & $+21.2$ {\scriptsize$[17.2, 26.0]$} & $+31.6$ {\scriptsize$[26.6, 38.0]$} \\
    Clip, $-10$\,dB    & 234 & 81.7  & $+24.4$ {\scriptsize$[16.2, 34.1]$} & $+34.0$ {\scriptsize$[26.1, 43.5]$} & 47.1 & $+24.3$ {\scriptsize$[19.1, 29.6]$} & $+55.3$ {\scriptsize$[46.5, 65.3]$} \\
    Clip, $-20$\,dB    & 234 & 116.0 & $-2.3$ {\scriptsize$[-8.2, 4.3]$}   & $+1.1$ {\scriptsize$[-3.5, 6.1]$}   & 70.4 & $+45.0$ {\scriptsize$[33.6, 58.4]$} & $+45.3$ {\scriptsize$[35.2, 58.6]$} \\
    4 words, $-10$\,dB & 223 & 26.2  & $+15.0$ {\scriptsize$[9.6, 22.7]$}  & $+22.4$ {\scriptsize$[18.6, 27.1]$} & 18.9 & $+4.7$ {\scriptsize$[2.9, 6.5]$}    & $+10.2$ {\scriptsize$[7.8, 12.9]$} \\
    4 words, $-20$\,dB & 223 & 39.1  & $+4.4$ {\scriptsize$[2.3, 6.7]$}    & $+9.9$ {\scriptsize$[6.9, 13.4]$}   & 25.4 & $+9.6$ {\scriptsize$[7.2, 12.4]$}   & $+13.5$ {\scriptsize$[10.6, 17.3]$} \\
    \bottomrule
  \end{tabular}
\end{table*}

\paragraph{The generator uses the lips.} If $G$ only spoke the text it is given, the lips it sees would not matter. We test this by changing only its video input. For each eligible sample, the replacement is the nearest nonoverlapping segment of the same source video that contains none of the target content words. Stop words and fillers are ignored for this exclusion. Both segments come from the same source video, and as an additional screen against other speakers we require a WavLM-Base-plus-SV speaker cosine of at least 0.86. The audio, the supplied transcript, the sampling seed and all inference settings are kept, and the replacement lips still move, so the intervention changes what the lips say and little else. A third input removes the articulation altogether: a frozen mouth, in which the first frame of the window is repeated over its whole length, as in the baseline of the visual evidence $e_R$. This third input lets us separate what the correct lips add from what conflicting lips cost. The experiment covers every sample with an eligible replacement: 1,877 of the 2,092 verified silenced cuts, and 914 severe-noise samples, in which babble at $-10$ or $-20$\,dB covers four words or the whole window. To measure the variation between two renders, we also generate 300 samples of each task again with their own video.

With the transcript held fixed, a frozen mouth lowers exact recovery at every span length and raises WER in every severe-noise condition, and mismatched lips cost more again (\cref{tab:supp:visual_intervention}). In these columns the transcript is still the one obtained from the correct lips, so the lip reader and the language model play no part in either loss, and both arise in the generator itself. The loss with a frozen mouth shows that the correct lips help the generator beyond the words it is given. The additional loss with mismatched lips shows that it follows those words less faithfully when the lips contradict them. Over all cuts and over all severe-noise samples, mismatched lips are also significantly worse than the frozen mouth. The effect of the lips is largest under severe noise. When babble at $-20$\,dB covers the whole clip, the generator voices the supplied transcript only with the correct video, and fails with a frozen mouth as completely as with mismatched lips. Two renders of the same sample with its own video differ by at most 0.1 points, so none of these differences is sampling variation.

Without a transcript, the generator depends on the lips wherever it has something to recover. A frozen mouth halves its recovery on short spans, and mismatched lips remove more. Longer spans are near zero even with the correct video, and whole-window $-20$\,dB babble is already beyond $G$ alone, so neither can fall further. When the lip reader also reads the mismatched lips, the loss under severe noise grows again ($+40.9$ points over all samples, CI $[35.0, 48.6]$). The visual signal therefore enters the pipeline twice, through recognition and through generation.

\subsection{Scoring checks}
\label{sec:supp:scoring_checks}

We check whether the comparisons depend on the silenced-cut verification filter or on the ASR model used to score generated speech. The outputs and selection choices in the second-ASR comparison are held fixed.

\begin{table}[htbp]
  \centering
  \small
  \setlength{\tabcolsep}{5.0pt}
  \caption{\textbf{Silenced speech without the verification filter.} Exact span recovery (\%) on all 2,328 TalkVid and VoxMM cuts, and our method minus each baseline with its 95\% interval. Bold: the interval excludes zero. \Cref{tab:supp:ci}a gives the same comparison on the 2,092 verified cuts.}
  \begin{tabular}{@{}lcc@{}}
    \toprule
    Method & Exact & $\Delta$ {\scriptsize[95\% CI]} \\
    \midrule
    Ours (Auto-AVSR as $R$) & 35.6 & \\
    \midrule
    Auto-AVSR          & 28.0 & $\mathbf{+7.6}$ {\scriptsize$[5.5, 9.9]$} \\
    USR\,2.0           & 32.7 & $+2.9$ {\scriptsize$[-0.9, 7.0]$} \\
    Llama-AVSR         & 28.1 & $\mathbf{+7.5}$ {\scriptsize$[2.3, 12.8]$} \\
    LipVoicer          & 25.8 & $\mathbf{+9.8}$ {\scriptsize$[6.0, 14.4]$} \\
    VoiceCraft ($L$)   & 29.2 & $\mathbf{+6.4}$ {\scriptsize$[5.0, 8.0]$} \\
    F5-TTS ($L$)       & 32.3 & $\mathbf{+3.3}$ {\scriptsize$[2.2, 4.3]$} \\
    AlignDiT ($L$)     & 32.2 & $\mathbf{+3.4}$ {\scriptsize$[2.2, 4.7]$} \\
    \bottomrule
  \end{tabular}
  \label{tab:supp:allcuts}
\end{table}

\paragraph{All cuts.} The primary analysis uses verified cuts because Whisper's errors on clean audio would otherwise confound restoration scoring (\cref{sec:supp:scoring}). As a sensitivity check, \cref{tab:supp:allcuts} scores all cuts except the 54 with incorrect reference transcripts. Every method's score falls, but differences from the baselines remain within a point and a half of those on verified cuts, with the same significant comparisons.

\paragraph{Independent ASR scoring.} We re-score the same outputs with Parakeet in place of Whisper, holding generated audio and selection decisions fixed. On noisy speech, the 15 methods in the original comparison retain their ranking, and ours remains first (\cref{tab:supp:parakeet_noisy}). Five of these methods are recognizers whose text is scored directly. Their scores are identical under both transcribers by construction, and the check is therefore informative for the ten methods whose output is speech. Among those, the lead of our selection over the external baselines persists. Its lead over fixed $R \rightarrow G$ is not significant under Parakeet, while the gain over fixed $R \rightarrow L \rightarrow G$ remains significant.

\begin{table}[htbp]
  \centering
  \scriptsize
  \setlength{\tabcolsep}{2pt}
  \caption{\textbf{Independent ASR check on noisy speech.} WER (\%) on 1,365 windows under Whisper and Parakeet, with the generated outputs and selection choices held fixed. $\Delta$: ours minus the method under Parakeet, with its 95\% interval. $^\dagger$Compared on the windows the method covers.}
  \begin{tabular}{@{}lrrl@{}}
    \toprule
    Method & Whisper & Parakeet & $\Delta$ [95\% CI] \\
    \midrule
    Ours, selection                 & \textbf{27.2} & \textbf{28.8} & -- \\
    $R \rightarrow G$              & 29.0 & 29.3 & $-0.5$ $[-1.0, 0.0]$ \\
    $R \rightarrow L \rightarrow G$ & 29.5 & 29.7 & $-0.9$ $[-1.5, -0.4]$ \\
    VisualVoice$^\dagger$           & 40.1 & 39.3 & $-10.4$ $[-13.4, -7.7]$ \\
    USR\,2.0 (V)                    & 47.4 & 47.4 & $-18.6$ $[-23.7, -14.0]$ \\
    AlignDiT$^\dagger$              & 50.5 & 50.6 & $-21.9$ $[-26.5, -17.2]$ \\
    LipVoicer                       & 52.9 & 54.0 & $-25.1$ $[-29.8, -20.3]$ \\
    Noisy input                     & 69.0 & 62.5 & $-33.7$ $[-37.7, -30.3]$ \\
    DeepFilterNet3                  & 72.2 & 74.6 & $-45.8$ $[-50.4, -41.8]$ \\
    Sidon                           & 84.4 & 85.1 & $-56.3$ $[-60.5, -52.8]$ \\
    SAM Audio, text prompt          & 67.1 & 67.2 & $-38.4$ $[-43.0, -34.2]$ \\
    SAM Audio, visual prompt        & 75.4 & 75.2 & $-46.4$ $[-50.9, -41.6]$ \\
    \bottomrule
  \end{tabular}
  \label{tab:supp:parakeet_noisy}
\end{table}

On the condition grid, Parakeet confirms a gain over a task router that uses $R \rightarrow L \rightarrow G$ for silenced speech and $R \rightarrow G$ otherwise (\cref{tab:supp:parakeet_grid_silenced}a, with the grid under Whisper in \cref{tab:supp:grid}). Under whole-clip $-20$\,dB babble, selection mutes the input on 236 of 268 windows and markedly outperforms that router. Under Parakeet it is statistically tied with the retrospective condition-wise rule of \cref{sec:supp:selection}. As noted there, that rule uses the test outcomes of each condition, but it is not the per-sample oracle of \cref{tab:supp:selection_ablation}.

\begin{table}[htbp]
  \centering
  \scriptsize
  \caption{\textbf{Independent ASR checks on variable corruption and silenced speech.} (a) Grid WER under Parakeet. $\Delta$: selection minus the comparator. The last two rows are the whole-clip $-20$\,dB babble condition. (b) Exact span recovery (\%) with every speech output transcribed by Parakeet, on the cuts verified from clean audio by Whisper or by Parakeet. Speech outputs therefore score differently from \cref{tab:supp:ci}a.}
  \setlength{\tabcolsep}{2pt}
  \textbf{(a) Condition grid (3,419 samples)}\\[2pt]
  \begin{tabular}{@{}lrl@{}}
      \toprule
      Rule & WER & $\Delta$ [95\% CI] \\
      \midrule
      Selection                 & \textbf{24.2} & -- \\
      Task router               & 26.3 & $-2.1$ $[-2.5, -1.8]$ \\
      Always $R \rightarrow G$  & 26.4 & $-2.2$ $[-2.6, -1.8]$ \\
      Always $R \rightarrow L \rightarrow G$ & 26.6 & $-2.5$ $[-2.8, -2.1]$ \\
      Condition-wise hindsight  & 24.3 & $-0.2$ $[-0.4, +0.1]$ \\
      \midrule
      Selection, $-20$\,dB babble & 51.1 & -- \\
      Task router, $-20$\,dB babble & 69.5 & $-18.4$ $[-21.8, -15.3]$ \\
      \bottomrule
  \end{tabular}\\[5pt]
  \textbf{(b) Silenced cuts, Parakeet scoring}\\[2pt]
  \begin{tabular}{@{}lrr@{}}
      \toprule
      Method & Whisper-verified (2,092) & Parakeet-verified (2,066) \\
      \midrule
      Ours, selection & \textbf{38.3} & \textbf{38.9} \\
      F5-TTS          & 35.8 & 36.1 \\
      USR\,2.0        & 35.8 & 36.3 \\
      AlignDiT        & 35.4 & 35.9 \\
      VoiceCraft      & 31.8 & 32.0 \\
      Auto-AVSR       & 30.7 & 30.9 \\
      Llama-AVSR      & 30.4 & 30.9 \\
      LipVoicer       & 27.2 & 27.3 \\
      \bottomrule
  \end{tabular}
  \label{tab:supp:parakeet_grid_silenced}
\end{table}

On silenced speech, Parakeet scoring preserves the relative significance pattern on both verified cut sets (\cref{tab:supp:parakeet_grid_silenced}b). In particular, USR\,2.0 remains statistically tied with our selection (difference $+2.5$ points, CI $[-1.7, +7.0]$ on Parakeet-verified cuts). The text-only baselines' scores change between columns because the verified cuts change, not because their outputs are re-transcribed.

\paragraph{Agreement with listeners.} The listening study of \cref{sec:supp:listening} provides a check that does not depend on any speech recognizer. The listeners' preferences on the words follow the WER differences reported here, with a Spearman correlation of 0.71 over the 120 windows of the study.

\paragraph{Clean input.} We also test whether restoration changes intact speech. Selection keeps the input audio for every clean clip on the condition grid (\cref{sec:supp:selection}). Across these 268 windows, its output has 10.1\% WER, compared with 9.9\% for the original recordings. The 0.2-point difference is not significant (CI $[-0.6, 1.0]$), indicating no measurable loss of intelligibility on this clean set.

\subsection{Illustrative examples}

\begin{figure*}[t]
  \centering
  \includegraphics[width=\linewidth]{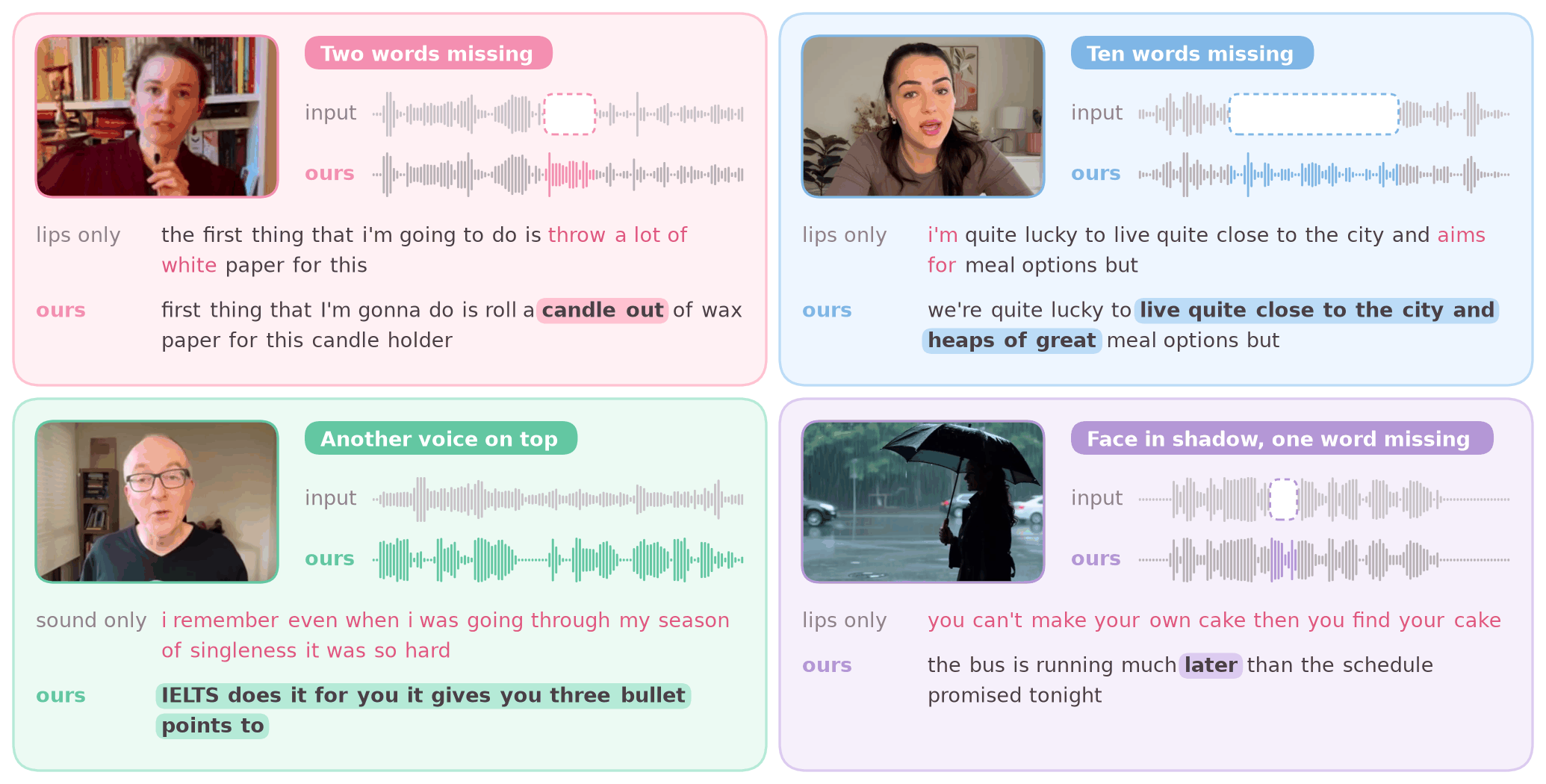}
  \caption{\textbf{Four test examples.} Each card shows a video frame, the loudness of the input and of our output ($R \rightarrow L \rightarrow G$), and two transcripts. Dashed boxes mark the silenced interval. \emph{Lips only} is the lip reader's best hypothesis and \emph{sound only} is Whisper's transcript of the input, with wrong words in red. \emph{Ours} is Whisper's transcript of our output, with the restored words highlighted.}
  \label{fig:supp:showcase}
\end{figure*}

\Cref{fig:supp:showcase} shows four test examples: two silenced spans of two and ten words, a competing speaker, and a speaker of the hard set whose face is in shadow. In each of them the lips alone or the sound alone give the wrong words, and our output is transcribed as the reference. \Cref{fig:supp:examples} shows the first and the third as spectrograms.

In the silenced example, the generated span joins the audible speech on both sides without a visible seam, and the audible parts are preserved. For the same cut, the generator alone and the generator with the lip reader's hypothesis produce speech that is transcribed as ``keletor of wax paper'' and ``gullet of wax paper''. Only with the transcript of the language model does it say ``candle out of wax paper''. In the noisy example the competing speaker dominates the input, to the point that Whisper transcribes their sentence and not the target's. The output removes that speaker and follows the timing of the clean recording, including its pause.

\subsection{Gaps without digital silence}
\label{sec:supp:nonzero_gaps}

In the main evaluation the removed span is digital silence, which the amplitude threshold of \cref{sec:supp:models} locates almost perfectly. Here we test whether the pipeline depends on the gap being digital silence. Using the same mute envelope, we fill the span with one of three kinds of audio. Room tone is assembled from the quietest 5\% of 32\,ms frames outside the span, with 5\,ms crossfades. Comfort noise is pink noise at the 10th percentile of the frame root-mean-square level outside the span. Attenuated speech is the original speech lowered by 40\,dB, the easiest case, since faint speech remains in the input by construction. All three fillers are synthetic corruptions of the existing test windows, not new recordings.

\paragraph{Detector.} We define a gap as an interval in which the audio contains no speech while the mouth keeps articulating. Silero VAD v5 proposes interior non-speech runs of at least 0.256\,s at threshold 0.7. A visual articulation score, $1-P(\text{blank})-P(\text{silence})$ from a viseme head on AV-HuBERT features, is smoothed over seven frames. The detector chooses the candidate with the greatest mean articulation and declares a gap only when that mean is at least 0.4. As an ablation, an audio-only detector takes the longest non-speech run (threshold 0.2, at least 0.448\,s). All parameters were selected on 373 validation cuts and 99 undamaged validation windows, to maximize mean detection across the three fillers subject to at most 5\% false alarms on undamaged windows, and were frozen before any test run. The chosen operating point has 4.0\% validation false alarms, and no audio-only setting met the limit (9.1\% at its chosen point). The true interval is used to construct the corruption and to score, never as an input.

\begin{table}[htbp]
  \centering
  \footnotesize
  \setlength{\tabcolsep}{3pt}
  \caption{\textbf{Localization of gaps without digital silence.} A gap is \emph{found} when the declared interval overlaps the true span with an intersection-over-union of at least 0.5. IoU and boundary error (\emph{Edge}) are medians over the declared intervals. Digital zero is a reference for this detector, which is separate from the amplitude threshold of the main evaluation.}
  \begin{tabular}{@{}llrrr@{}}
    \toprule
    Set & Filler & Found (\%) & Median IoU & Edge (ms) \\
    \midrule
    TalkVid & Digital zero      & 91.2 & 0.966 & 13 \\
            & Room tone         & 82.3 & 0.880 & 61 \\
            & Comfort noise     & 84.6 & 0.859 & 70 \\
            & Attenuated speech & 89.6 & 0.953 & 20 \\
    \midrule
    VoxMM   & Digital zero      & 90.0 & 0.966 & 16 \\
            & Room tone         & 84.2 & 0.900 & 56 \\
            & Comfort noise     & 84.4 & 0.882 & 70 \\
            & Attenuated speech & 85.1 & 0.945 & 20 \\
    \bottomrule
  \end{tabular}
  \label{tab:supp:nonzero_gap_detection}
\end{table}

\paragraph{Localization.} On the verified test cuts, the detector finds more than four in five of these gaps on both sets, with median boundary errors of at most 70\,ms (\cref{tab:supp:nonzero_gap_detection}). On undamaged windows it declares a gap in 2.7\% of 113 TalkVid windows and 9.2\% of 153 VoxMM windows.

\begin{table}[htbp]
  \centering
  \scriptsize
  \setlength{\tabcolsep}{2.6pt}
  \caption{\textbf{Recovery with detected gaps.} Exact span recovery (\%) of the language model's transcript on the 2,092 verified cuts. \emph{Decl.}: share of cuts (\%) with a declared gap. A missed gap counts as not recovered. $\Delta$: difference from the true interval, with its 95\% interval. \emph{On decl.}: recovery on the cuts with a declared gap, using the detected interval and the true one. The first block moves the true interval by 100\,ms.}
  \begin{tabular}{@{}llrrlr@{}}
    \toprule
    Gap & Detector & Decl. & Exact & $\Delta$ {\scriptsize[95\% CI]} & On decl. \\
    \midrule
    Digital zero   & true interval & --   & 36.1 & --                                        & -- \\
                   & edges outward & --   & 35.6 & $-0.5$ {\scriptsize$[-1.5,+0.5]$}         & -- \\
                   & edges inward  & --   & 36.1 & $+0.0$ {\scriptsize$[-0.7,+0.8]$}         & -- \\
                   & shifted later & --   & 36.6 & $+0.5$ {\scriptsize$[-0.3,+1.2]$}         & -- \\
    \midrule
    Room tone      & audio + lips  & 88.1 & 30.4 & $-5.6$ {\scriptsize$[-8.0,-3.8]$}         & 34.6 / 34.6 \\
    Comfort noise  & audio + lips  & 87.9 & 29.6 & $-6.5$ {\scriptsize$[-8.9,-4.6]$}         & 33.7 / 34.2 \\
    Attenuated     & audio + lips  & 92.0 & 33.2 & $-2.9$ {\scriptsize$[-5.1,-1.1]$}         & 36.1 / 35.8 \\
    \midrule
    Room tone      & audio         & 76.7 & 23.8 & $-12.3$ {\scriptsize$[-15.5,-9.6]$}       & 31.0 / 31.1 \\
    Comfort noise  & audio         & 79.9 & 24.8 & $-11.3$ {\scriptsize$[-13.9,-8.9]$}       & 31.1 / 31.2 \\
    Attenuated     & audio         & 85.7 & 29.2 & $-6.9$ {\scriptsize$[-8.7,-5.0]$}         & 34.0 / 33.6 \\
    \bottomrule
  \end{tabular}
  \label{tab:supp:nonzero_gap_recovery}
\end{table}

\paragraph{Recovery.} We then build the audible context from the detected interval and run the language model of the main experiments, with no other change. Since the detected interval is used only to build this context, we score the transcript that the language model returns. With the true interval, this reproduces the $R \rightarrow L$ result of \cref{tab:ablations}a to within 0.1 points. A missed gap is scored as not recovered, which makes these numbers a lower bound.

Whenever a gap is declared, recovery equals that obtained with the true interval, to within half a point for every filler and both detectors (\cref{tab:supp:nonzero_gap_recovery}, last column). The loss relative to the true interval therefore comes from the missed gaps and not from imprecise boundaries. Consistently, moving both true boundaries 100\,ms outward or inward, or shifting the interval by 100\,ms, leaves recovery unchanged to within half a point, and the detector's median boundary error is smaller than that. Placing the gap therefore does not rely on digital silence.

The lips also help to find the gap. On the same cuts, the detector that reads them recovers 4 to 7 points more spans than the audio-only one, and every interval excludes zero.

\subsection{What the language model takes from the lip reader}
\label{sec:supp:lm_hypotheses}

The language model sees the audible context and the lip reader's $N$-best list. To separate the two, we change only the list it is shown and keep everything else of the $R \rightarrow L$ row of \cref{tab:ablations}a: the cuts, the context, the gap duration and the decoding. We remove the list, replace it with the list of a window from another source video whose top hypothesis has a similar number of words, or shorten it to its first one or five hypotheses (\cref{tab:supp:lm_hypotheses}).

\begin{table}[htbp]
  \centering
  \footnotesize
  \setlength{\tabcolsep}{1pt}
  \caption{\textbf{Hypotheses shown to the language model.} Exact recovery (\%) of the $R \rightarrow L$ transcript on the verified silenced cuts. $\Delta$: difference to the full list, in points, with its 95\% interval, bold where the interval excludes zero.}
  \label{tab:supp:lm_hypotheses}
  \begin{tabular*}{\columnwidth}{@{}l@{\extracolsep{\fill}}r@{\extracolsep{0pt}\hspace{4pt}}l@{\extracolsep{\fill}}r@{\extracolsep{0pt}\hspace{4pt}}l@{}}
    \toprule
    & \multicolumn{2}{c}{TalkVid + VoxMM} & \multicolumn{2}{c}{Hard} \\
    \cmidrule{2-3} \cmidrule{4-5}
    Shown to $L$ & \% & $\Delta$ {\scriptsize[95\% CI]} & \% & $\Delta$ {\scriptsize[95\% CI]} \\
    \midrule
    Context only           & 3.2  & $\mathbf{-33.0}$ {\scriptsize$[-43.7, -25.0]$} & 23.3 & $+0.9$ {\scriptsize$[-14.0, 13.9]$} \\
    Other video's          & 2.3  & $\mathbf{-33.9}$ {\scriptsize$[-44.4, -24.9]$} & 17.2 & $-5.2$ {\scriptsize$[-17.9, 5.2]$} \\
    $h_1$ only             & 32.0 & $\mathbf{-4.2}$ {\scriptsize$[-6.5, -2.1]$}    & 22.4 & $0.0$ {\scriptsize$[-5.2, 5.0]$} \\
    First five             & 35.0 & $-1.1$ {\scriptsize$[-2.4, 0.0]$}              & 25.9 & $+3.4$ {\scriptsize$[0.0, 7.9]$} \\
    All ten (ours)         & 36.2 &                                                & 22.4 & \\
    \midrule
    Lip reader             & 30.7 & $\mathbf{-5.5}$ {\scriptsize$[-7.6, -3.3]$}    & 8.6  & $\mathbf{-13.8}$ {\scriptsize$[-21.7, -5.1]$} \\
    \bottomrule
  \end{tabular*}
\end{table}

On TalkVid and VoxMM, the language model cannot fill the span without the lip reader. With the context alone, or with the hypotheses of another video, it recovers almost nothing, far below the lip reader itself. It does not copy the top hypothesis either: with $h_1$ alone it stays close to the lip reader, and its gain grows with the list, most of it reached with five hypotheses. Of the spans it recovers with the full list, 13\% are correct in a lower-ranked hypothesis but not in $h_1$, and 9\% in none of the ten. The language model thus treats the list as candidates, which it chooses among and edits using the context.

The hard set behaves differently, as expected from its construction. Its mouths are difficult or impossible to read, so the hypotheses carry little information: the language model recovers as many spans from the context alone as with the full list, and 65\% of the spans it recovers appear in none of the ten hypotheses. The same context-only input that fails on TalkVid and VoxMM thus works where the context is informative, so the drop above reflects the missing lip information and not the form of the input. This matches the behaviour reported in \cref{sec:exp:hard}: when the lips cannot be read, the context compensates.

\section{Speech quality, speaker similarity and lip synchronization}
\label{sec:supp:quality}

\Cref{tab:supp:quality} complements the word-level results with the quality of the generated speech. We measure naturalness with DNSMOS~\cite{reddy2021dnsmos} and UTMOS~\cite{saeki2022utmos}, lip synchronization with LSE-C and LSE-D~\cite{prajwal2020lip,chung2016out}, and speaker similarity as the cosine similarity of ERes2Net~\cite{chen2023enhanced} speaker embeddings between the output and the clean recording of the same window. ERes2Net is the speaker model used by Just-Dub-It~\cite{Chen_2026}.

We prefer it to WavLM speaker embeddings because it separates speakers far more clearly on our data. On pairs of clean test recordings, two windows of the same video reach a mean ERes2Net similarity of 0.61 and windows of different videos 0.05, whereas WavLM gives 0.89 and 0.59, and 5\% of the different-speaker pairs score above 0.87 with WavLM. A WavLM similarity near 0.9 therefore does not show that two recordings share a voice.

\begin{table}[htbp]
  \centering
  \footnotesize
  \setlength{\tabcolsep}{2.6pt}
  \caption{\textbf{Quality of the generated speech}, pooled over TalkVid and VoxMM: naturalness (DNSMOS and UTMOS, on a scale from 1 to 5), speaker similarity to the clean recording (Spk.), and lip synchronization (LSE-C, LSE-D). For silenced speech, Spk.\ is measured on the silenced span alone. Bold and underline mark the best and second best method. $^\dagger$Only windows with successful preprocessing.}
  \begin{tabular}{@{}lrrrrr@{}}
    \toprule
    Method & DNSMOS$\uparrow$ & UTMOS$\uparrow$ & Spk.$\uparrow$ & LSE-C$\uparrow$ & LSE-D$\downarrow$ \\
    \midrule
    \multicolumn{6}{@{}l}{\emph{Silenced speech}} \\
    Clean recording                              & 2.54 & 2.77 & 0.839 & 6.38 & 8.07 \\
    Input (span silenced)                        & 2.58 & 2.62 & --    & 4.07 & 10.39 \\
    LipVoicer~\cite{yemini2024lipvoicer}         & 2.60 & 2.32 & 0.102 & 5.10 & 9.49 \\
    F5-TTS~\cite{chen-etal-2025-f5} ($L$)        & \textbf{2.71} & 2.54 & 0.597 & 4.70 & 9.70 \\
    AlignDiT~\cite{Choi_2025} ($L$)              & 2.57 & 2.62 & 0.528 & 5.49 & 8.95 \\
    Ours, $R \rightarrow G$                      & 2.69 & 2.68 & \underline{0.624} & \textbf{6.21} & \textbf{8.32} \\
    Ours, $R \rightarrow L \rightarrow G$        & \underline{2.70} & \underline{2.69} & \textbf{0.629} & 6.16 & 8.36 \\
    Ours, selection                              & \textbf{2.71} & \textbf{2.71} & 0.616 & \underline{6.20} & \underline{8.33} \\
    \midrule
    \multicolumn{6}{@{}l}{\emph{Noisy speech}} \\
    Clean recording                              & 2.57 & 2.78 & --    & 6.38 & 8.05 \\
    Noisy input                                  & 1.33 & 1.38 & 0.492 & 2.36 & 11.02 \\
    DeepFilterNet3~\cite{schroter23b_interspeech} & 2.14 & 1.54 & 0.382 & 3.07 & 10.93 \\
    Sidon~\cite{nakata2026sidon}                 & 2.47 & 2.60 & 0.185 & 2.26 & 12.37 \\
    VisualVoice~\cite{Gao_2021_CVPR}$^\dagger$   & 1.63 & 1.46 & 0.654 & 4.35 & 9.30 \\
    LipVoicer~\cite{yemini2024lipvoicer}         & 2.60 & 2.32 & 0.124 & 5.08 & 9.48 \\
    AlignDiT~\cite{Choi_2025} ($R$)$^\dagger$    & \underline{2.75} & \textbf{3.10} & 0.525 & \underline{6.31} & 8.27 \\
    Ours, $G$ alone                              & \textbf{2.82} & \underline{2.70} & \underline{0.664} & \textbf{6.33} & \underline{8.17} \\
    Ours, $R \rightarrow G$                      & 2.61 & 2.58 & \textbf{0.698} & 6.30 & \textbf{8.15} \\
    Ours, selection                              & 2.67 & \underline{2.70} & 0.660 & 6.24 & 8.28 \\
    \bottomrule
  \end{tabular}
  \label{tab:supp:quality}
\end{table}

\begin{table*}[t]
  \centering
  \footnotesize
  \setlength{\tabcolsep}{2.4pt}
  \caption{\textbf{Listening study on noisy speech.} Share of judgments (\%) that prefer our output (\emph{Ours}), rate the two the same, or prefer the baseline (\emph{Base.}), from 16 raters and 64 judgments per baseline and question. \emph{Score}: mean preference from $-1$ (baseline) to $+1$ (ours) with its 95\% interval over raters, bold where it excludes zero.}
  \label{tab:supp:listening}
  \begin{tabular}{@{}l @{\hspace{9pt}} rrrrl @{\hspace{9pt}} rrrrl @{\hspace{9pt}} rrrrl@{}}
    \toprule
    & \multicolumn{5}{c}{Same words} & \multicolumn{5}{c}{Same speaker} & \multicolumn{5}{c}{Cleaner and more natural} \\
    \cmidrule(r{6pt}){2-6} \cmidrule(r{6pt}){7-11} \cmidrule{12-16}
    Ours vs. & Ours & Same & Base. & \multicolumn{2}{c}{Score {\scriptsize[95\% CI]}} & Ours & Same & Base. & \multicolumn{2}{c}{Score {\scriptsize[95\% CI]}} & Ours & Same & Base. & \multicolumn{2}{c}{Score {\scriptsize[95\% CI]}} \\
    \midrule
    DeepFilterNet3   & 81.3 & 15.6 & 3.1 & $\mathbf{+0.78}$ & {\scriptsize$[0.66, 0.89]$} & 87.5 & 10.9 & 1.6 & $\mathbf{+0.86}$ & {\scriptsize$[0.73, 0.97]$} & 92.2 & 6.3 & 1.6 & $\mathbf{+0.91}$ & {\scriptsize$[0.81, 0.98]$} \\
    Sidon            & 84.4 & 12.5 & 3.1 & $\mathbf{+0.81}$ & {\scriptsize$[0.72, 0.91]$} & 90.6 & 7.8 & 1.6 & $\mathbf{+0.89}$ & {\scriptsize$[0.78, 0.97]$} & 93.8 & 4.7 & 1.6 & $\mathbf{+0.92}$ & {\scriptsize$[0.84, 0.98]$} \\
    VisualVoice      & 48.4 & 35.9 & 15.6 & $\mathbf{+0.33}$ & {\scriptsize$[0.16, 0.52]$} & 60.9 & 29.7 & 9.4 & $\mathbf{+0.52}$ & {\scriptsize$[0.28, 0.73]$} & 82.8 & 14.1 & 3.1 & $\mathbf{+0.80}$ & {\scriptsize$[0.69, 0.91]$} \\
    LipVoicer        & 78.1 & 17.2 & 4.7 & $\mathbf{+0.73}$ & {\scriptsize$[0.50, 0.89]$} & 92.2 & 6.3 & 1.6 & $\mathbf{+0.91}$ & {\scriptsize$[0.81, 0.98]$} & 90.6 & 7.8 & 1.6 & $\mathbf{+0.89}$ & {\scriptsize$[0.80, 0.97]$} \\
    AlignDiT ($R$)   & 53.1 & 31.3 & 15.6 & $\mathbf{+0.38}$ & {\scriptsize$[0.19, 0.56]$} & 46.9 & 23.4 & 29.7 & $+0.17$ & {\scriptsize$[-0.02, 0.38]$} & 45.3 & 21.9 & 32.8 & $+0.13$ & {\scriptsize$[-0.05, 0.28]$} \\
    \midrule
    All baselines    & 69.1 & 22.5 & 8.4 & $\mathbf{+0.61}$ & {\scriptsize$[0.52, 0.68]$} & 75.6 & 15.6 & 8.8 & $\mathbf{+0.67}$ & {\scriptsize$[0.58, 0.74]$} & 80.9 & 10.9 & 8.1 & $\mathbf{+0.73}$ & {\scriptsize$[0.68, 0.78]$} \\
    \bottomrule
  \end{tabular}
\end{table*}

For silenced speech, methods differ chiefly around the missing span, so we measure speaker similarity on that span, for the 1,586 cuts whose span lasts at least 1\,s. Over the whole window it is nearly identical for F5-TTS and our method (0.878 and 0.879), since most of each window is audible speech that both reproduce. On the span, each of our rows is above both editors, significantly so (CI $[0.009, 0.029]$ for the selection against F5-TTS, the closest pair). As a reference for the scale of this measure, the clean row compares the span of the clean recording with its own window. Our outputs are also the closest to the clean recordings in lip synchronization, well ahead of the editors, and UTMOS places them above the editors and LipVoicer and just below the clean recordings. DNSMOS rates most generated outputs above even the clean recordings, in both tasks, so we read it together with UTMOS.

On noisy speech, the highest DNSMOS and lip synchronization scores are those of our fixed configurations $G$ alone and $R \rightarrow G$. The selection and AlignDiT follow closely, and AlignDiT is the only method with a higher UTMOS. These scores rate how natural and well synchronized the speech sounds, and none of them measures whether its words are correct. AlignDiT speaks the lip reader's hypothesis in a clean reference voice, which UTMOS rates above the clean recordings themselves, but its WER is 23.4 points higher than that of the selection on the same windows (\cref{tab:supp:ci}b). It also discards the noisy input, and its speaker similarity is below that of every one of our rows. The selection is slightly below $R \rightarrow G$ in speaker similarity. This difference is concentrated in the windows where the selection mutes the input: similarity there is 0.191, compared with 0.727 when the input audio is kept. In those windows no usable voice remains in the input, and the generator infers one from the speaker's appearance. The result is above the similarity of different speakers (0.05) and above LipVoicer, which also generates from the video alone, but well below the windows that keep the audio. Muting thus trades the recorded voice for intelligibility.

\subsection{Listening study}
\label{sec:supp:listening}

All results so far rely on automatic metrics. To check that they agree with human perception, we also run a listening study on the noisy-speech task, in which raters compare the output of our selection with that of one baseline at a time.

\paragraph{Protocol.} Each page presents one noisy test window as three clips with the same video: the clean recording as the reference, and two restorations, ours and that of one baseline. Their order is balanced across pages and is not revealed to the rater. The rater answers three questions: ``Which one says the same words as the original?'', ``Which one sounds more like the same person?'' and ``Which one sounds cleaner and more natural?''. Each answer is one of the two restorations or ``about the same''.

The study covers 120 windows of TalkVid and VoxMM, 24 for each of five baselines, across the noise conditions of \cref{tab:noisy} and the babble sources of \cref{sec:supp:babble}. On these windows our output has 30.7\% WER, slightly above its 27.2\% on the full noisy set, so the sample is not easier for our method than the evaluation as a whole. Sixteen raters each judged 20 windows, and all of them passed two attention checks. This gives 960 judgments, and the intervals in \cref{tab:supp:listening} are obtained by resampling the raters. All raters gave their consent on the first page of the study, and their anonymous answers are used only to compute the statistics reported here. Because the study records no identifying information and collects only listening preferences, no ethics board review was needed.

\paragraph{Results.} Listeners prefer our output on all three questions (\cref{tab:supp:listening}), and each of the 16 raters prefers it on average. The preference is strongest against the two enhancement models and LipVoicer. VisualVoice is the closest baseline on the words, as it is in WER (\cref{tab:supp:ci}b). AlignDiT is the one baseline that listeners rate level with ours in voice and quality, where both intervals include zero. It is given a clean recording of the speaker (\cref{tab:supp:inputs}) and speaks in that voice, which the automatic scores of \cref{tab:supp:quality} also rate highly. On the words, listeners prefer our output over it as well.

The listeners' judgments also agree with the automatic scoring. Across the 120 windows, the mean preference on the words correlates with the WER difference between the baseline and our output (Spearman $\rho = 0.71$). Because these judgments do not depend on any speech recognizer, the agreement supports the Whisper-based scoring used throughout the paper.

\section{Limitations and responsible use}
\label{sec:supp:limitations}

\paragraph{Scope of the evaluation.} We evaluate on 4.8\,s windows with one damaged region each, and the context given to the language model is the audible part of the same window. Longer recordings can be processed window by window, but we do not study how context from earlier windows could be carried forward. The corruptions are synthetic, so that the hidden words are known and can be scored. \Cref{sec:supp:nonzero_gaps} extends them to gaps filled with room tone or comfort noise, but recordings damaged by packet loss, codecs or reverberation are not covered. Beyond English, we test Spanish and French on a smaller set (\cref{tab:ablations}b), and any other language needs a lip reader for that language.

\paragraph{Dependence on the components.} Our method inherits the limits of its components. When neither the audio nor the context carries usable information, as in a clip fully covered by loud babble, the result is bounded by what the lip reader can read (\cref{tab:supp:grid}). The selection can also follow a wrong lip reading when the lips are hard to read, as on the silenced hard set (\cref{sec:exp:ablations}). Since the lip reader is frozen and can be swapped, a stronger one improves the result without retraining (\cref{tab:silenced}).

\paragraph{Voice.} When the selection discards the recorded audio, no usable voice remains in the input, and the generator infers one from the speaker's appearance (\cref{sec:supp:quality}). The words are restored, but the voice is then not the one in the recording. A clean recording of the same speaker, when one exists, could serve as a voice reference for these inputs, which we leave to future work.

\paragraph{Cost.} The selection runs the generator six times per window (\cref{sec:supp:runtime}), so our method is suited to offline restoration and not to live use. A single fixed configuration, $R \rightarrow L \rightarrow G$, needs one pass and stays close to the selection on both tasks of the standard test sets (\cref{tab:ablations}a), which gives a cheaper operating point.

\paragraph{Responsible use.} Our method generates speech for a real person, and a restored word can be wrong: the output is a reconstruction, not a recording. We therefore see it as a tool for restoring damaged recordings, archival footage and calls with the consent of the people in them. Its output should be marked as restored wherever it is shown, with the restored interval indicated when it is known, and it should not be treated as evidence of what a person said. The text that the generator speaks is derived from the speaker's own lip movements and the audible context, since the aim is to recover what was said, not to produce new speech for a person.

\end{document}